\documentclass[a4paper,fleqn,usenatbib]{mnras}

\usepackage{newtxtext,newtxmath}
\usepackage{mathptmx}

\usepackage[T1]{fontenc}
\usepackage{ae,aecompl}

\usepackage{animate}

\usepackage{multirow,bigdelim}

\usepackage{graphicx}	

\title[Radio study of MACS clusters with GMRT]{Low-frequency radio study of MACS clusters at 610 and 235 MHz using the GMRT}
\author[Paul S.]{Surajit Paul$^{1,2}$\thanks{E-mail: (SP) surajit@physics.unipune.ac.in}, Sameer Salunkhe$^{1}$, Abhirup Datta$^{3}$ and Huib T. Intema $^{4,5}$\\
$^{1}$Department of Physics, Savitribai Phule Pune University, Pune 411007, India\\
$^{2}$Inter University Centre for Astronomy and Astrophysics, Pune 411007, India\\
$^3$Centre of Astronomy, Indian Institute of Technology Indore, Simrol, Khandwa Road, Indore 453552, India\\
$^4$Leiden Observatory, Leiden University, Niels Bohrweg 2, 2333 CA, Leiden, The Netherlands\\
$^5$International Centre for Radio Astronomy Research -- Curtin University, GPO Box U1987, Perth, WA 6845, Australia\\
}

\date{Accepted XXX. Received YYY; in original form ZZZ}

\pubyear{2015}

\begin{document}

\label{firstpage}
\pagerange{\pageref{firstpage}--\pageref{lastpage}}

\maketitle




%
%
%
%
%
%
%



\begin{abstract}

Studies have shown that mergers of massive galaxy clusters produce shocks and turbulence in the intra-cluster medium, the possible event that creates radio relics, as well as the radio halos. Here we present GMRT dual-band (235 and 610~MHz) radio observations of four such clusters from the MAssive Cluster Survey (MACS) catalogue. We report the discovery of a very faint, diffuse, elongated radio source with a projected size of about 0.5~Mpc in cluster MACSJ0152.5-2852. We also confirm the presence of a radio relic-like source (about 0.4~Mpc, previously reported at 325~MHz) in MACSJ0025.4-1222 cluster. Proposed relics in both these clusters are found apparently inside the virial radius instead of their usual peripheral location, while no radio halos are detected. These high-redshift clusters ($z=0.584$ and $0.413$) are among the earliest merging systems detected with cluster radio emissions. In MACSJ1931-2635 cluster, we found a radio mini-halo and an interesting highly bent pair of radio jets. Further, we present here a maiden study of low frequency (GMRT $235\;\&\;610$~MHz) spectral and morphological signatures of a previously known radio cluster MACSJ0014.3-3022 (Abell~2744). This cluster hosts a relatively flat spectrum ($\alpha^{610}_{235}\sim -1.15$), giant ($\sim 1.6$~Mpc each) halo-relic structure and a close-by high-speed ($1769\pm^{148}_{359}$~km~s$^{-1}$) merger-shock ($\mathcal{M}=2.02\pm^{0.17}_{0.41}$) originated from a possible second merger in the cluster.
\end{abstract}



\section{Introduction}\label{intro}

Clusters of galaxies are the largest ($\sim$ Megaparsec scale), gravitationally bound structures in the Universe. They are in general in thermal equilibrium and are prominent X-ray sources \citep{Sarazin1986}. But, a number of galaxy clusters are also detected at radio wavelengths (for review: \citealt{Weeren_2019SSRv}), confirming the presence of relativistic electrons and magnetic fields in the Intra-Cluster Medium (ICM), as well as the presence of strong dynamical activities in these systems \citep{Cuciti_2015A&A,Cassano_2010ApJ}. These radio sources are mostly diffuse in nature, connected to the ICM and not directly associated with the individual galaxies inside the clusters. They are usually steep spectrum ($\alpha <-1$) due to ageing effect \citep{Feretti_2012A&ARv}, or because of the nature of the injection spectra \citep{Stroe_2014MNRAS}. The large-scale ($l \gtrsim500$ kpc) diffuse radio sources are commonly divided into two broad classes, `radio halos'  and `radio relics' \citep{Feretti_2012A&ARv,Giovannini1999,Ensslin1998}.  Radio halos have a smooth morphology, are extended with sizes $\gtrsim$1 Mpc, steep spectrum, un-polarized and having low surface brightness ($\sim 0.1-1\mu$Jy arcsec$^{-2}$ at 1.4 GHz; \citealt{Giovannini_2009A&A,Feretti_2008LNP}), and are permeating the central volume of  clusters almost co-spatial with the thermal X-ray emitting gas of the ICM (e.g. Abell 2219, Abell 2163 etc. \citealt{Orr2007A&A,Feretti_2001A&A}). Giant radio relics (e.g., elongated arc-like), on the other hand, are mostly found at the periphery of the clusters, with sizes up to several Mpc, comparatively flatter spectrum and are highly polarized (p$\sim10-50\%$ at 1.4 GHz; \citealt{Weeren_2009A&A}). Discussion on another class of cluster radio sources has gained momentum in recent years, known as the radio mini-halos. Mini-halos are diffuse, low brightness, steep spectrum (i.e. $\alpha<-1$) radio structures found surrounding the Brightest Cluster Galaxies (BCGs) of the host clusters with typical extensions of $100-500$ kpc.

Radio synchrotron emission from galaxy clusters is a transient phenomenon ($\sim0.5$~Gyr) on cosmological time scales \citep{Cassano_2016A&A}, therefore should be linked to the dynamical states of the systems \citep{Donnert_2013MNRAS,Paul2012,Cassano2011,Rephaeli2008}.
Theoretically, in the large-scale structure formation framework, before attaining virialization, galaxy clusters pass through phases of continuous accretion and series of events of mergers of bigger and bigger galaxy groups \citep{Sarazin_2002ASSL}. Major cluster mergers are extremely energetic processes ($\sim$10$^{63-65}$~erg~s$^{-1}$, depending on merging masses and impact angle) that dissipate its energy in the ICM by thermalizing it through strong collision-less shocks \citep{Sarazin1986}. These shocks, by an efficient Fermi acceleration of ICM plasma, generate powerful MHD waves in the shocked medium. Also, shock compression significantly amplifies the upstream magnetic fields \citep{luigi2012,bykov2008}. A fraction of shock kinetic energy gets converted to turbulent energy and injects a volume-filling turbulence in the ICM \citep{Subra2006}. Particle acceleration due to Diffusive Shock Acceleration (DSA) and magnetic field amplification due to shock compression is most effective at the peripheral, high-Mach ($\mathcal{M} \gtrsim 3$) shock fronts \citep{luigi2012}, expected to produce the brightest radio synchrotron emission in form of relics \citep{Weeren_2010Sci}. Turbulent re-acceleration, on the other hand, expected to dominate at the central part of the clusters, where the level of merger-induced turbulent energy, availability of relativistic charged particles are sufficient to produce the radio halo emission \citep{Paul_2018arXiv,Pinzke2017MNRAS,Brunetti_2001MNRAS}.   But, mini-halos are not related to the cluster mergers, rather, they are only found in relaxed and massive systems with cool-cores and BCGs. Most probable mechanism is the re-acceleration of relic relativistic electron populations by MHD turbulence produced by AGN activity, cooling flow and core sloshing in cool-core clusters \citep{Gitti_2018A&A}.
 
As discussed above,  the cluster diffuse radio emissions are mostly connected to massive and active systems. In this study, objects are thus chosen from the MAssive Clusters Survey (MACS; \citealt{Ebeling2010,Horesh2010,Ebeling2001}) catalogue. The clusters are selected depending on their disturbed X-ray morphology and clumpy and elongated matter distribution as estimated from lensing studies to ensure its active dynamical phase. Four chosen clusters are Abell 2744, MACSJ0025.4-1222 (MACS0025), MACSJ0152.5-2852 (MACS0152) and MACSJ1931.8-2635 (MACS1931). Among them, cluster MACS1931 has a recently disturbed cool-core \citep{Ehlert_2011MNRAS}. At the time of these observations, only one  cluster i.e. Abell 2744, was known to host large-scale radio emission. A large halo and a relic were first reported at VLA 1.4 GHz \citep{Govoni2001A&A,Govoni2001} Later, in cluster MACS0025 a double relic has also been discovered at GMRT 325 MHz by \citet{Riseley2017A&A}. We performed a low-frequency radio spectral and morphological study at GMRT 235 and 610 MHz, as in general, diffuse radio emission from clusters is steep spectrum ($\alpha \le -1.0$) and expected to be better detected at low-frequencies.

We introduce our work in Section~\ref{intro}. Selection of clusters of galaxies, observation details and data analysis are described in Section~\ref{obs-detl}. In the result Section~\ref{result}, we report our findings from each observed objects. Section~\ref{upper-limit} is dedicated to calculation of halo upper limits. Thereafter, we discuss our findings in Section~\ref{disc} and finally, conclude the paper in Section~\ref{conc}. Cosmology used in this study is as follows: $H_0=70.2 \;\rm{km\;s^{-1}\;Mpc^{-1}}$, $\Omega_m=0.274$ and $\Omega_{\Lambda}=0.726$.

\section{Object Selection, GMRT observations and data analysis}\label{obs-detl}

\subsection{Selected objects}

\begin{table} 
\caption{Observation details} %
\label{tab:radioobs-list}      
\centering          
\begin{tabular}{|lclc|}     
\hline\hline       
Cluster name &  $z$&  Obs. date & Obs. time \\
\hline  
MACSJ0014.3-3022  &  0.308$^{a}$ &  13-AUG-2011 & 05:31 hour \\
(Abell 2744) & & 24-NOV-2006 & 04:42 hour${\dagger}$\\
MACSJ0025.4-1222  & 0.584$^{b}$ & 02-JUN-2011  & 03:43 hour \\
& &\& 05-JUL-2011&\\
& & 12-JAN-2014& 03:40 hour$\dagger$\\
MACSJ0152.5-2852  &  0.413$^{b}$ & 05-JUL-2011 & 04:50 hour \\ 
MACSJ1931.8-2635 &  0.352$^{a}$ &  03-JUN-2011 & 04:59 hour \\ 
\hline
\hline
\multicolumn{4}{l}{Col. 1: cluster name }\\ 
\multicolumn{4}{l}{Col. 2: redshift ($a:$\citet{Ebeling2010}, $b:$ \citet{Horesh2010},}\\
\multicolumn{4}{l}{Col. 3: GMRT Observation date and} \\
\multicolumn{4}{l}{Col. 4: On-source observation time}\\
\multicolumn{4}{l}{$\dagger$ This is an archival GMRT data at 325 MHz }\\
\end{tabular}
\end{table}

For this study, four massive galaxy clusters (see Table~\ref{tab:radioobs-list}) were chosen from the MACS catalogue. The MACS cluster survey was designed to find the population of strongly evolving clusters, with the most X-ray luminous systems using a specific X-ray selection function described in \citep{Ebeling2001}. From the MACS catalogues (up to 2010; \citealt{Ebeling_2007ApJ,Ebeling2010}) we selected clusters that show dynamical activities at different phases of mergers as indicated in X-ray/temperature maps, as well as weak lensing mass distribution (See \citealt{Zitrin2011}).

\subsection{GMRT observations}

Observations were done with the Giant Metre-wave Radio Telescope (GMRT) array, using 235 \& 610~MHz dual-band during June-August 2011 (Project Code : 20$\_$062). Observation details are provided in Table~\ref{tab:radioobs-list}. In this mode of GMRT observations, both frequencies are observed simultaneously in one polarization (Stokes RR at 610 MHz and LL at 235 MHz). Although, the observations were recorded over 32~MHz bandwidths at both frequencies, for 235~MHz observations, only a part of that bandwidth is available due to use of a bandpass filter. Each of the sources was observed with 4-5 hours of on-source time. Special care was taken to preserve maximum possible short baselines during the observations to properly image most of the diffuse radio emissions. Along with this, GMRT archival data of cluster Abell 2744 and MACS0025 were also analysed, both at 325 MHz.

\subsection{Data analysis}

\begin{table} 
\caption{Data analysis details} %
\label{tab:data-anal}      
\centering          
\begin{tabular}{|lccc|}     
\hline\hline       
Cluster name & Frequency & Robust &  UV taper \\
& (MHz) & & (Kilo-lambda) \\
\hline
Abell 2744  & 235 \& 610 & -1.0 & 10\\
& 325$^{\ast}$ & -1.0 & 0\\
MACS0025  & 235 \& 610 & 0.5  & 0\\
& 325$^{\dagger}$ & 0.5 & 0\\
MACS0152  & 235 \& 610 &
-1.0 & 10\\  
MACS1931 & 235 \& 610 & 0.0\&0.0 & 15\&20\\ 

\hline
\hline
\multicolumn{4}{l}{Col. 1: cluster name }\\ 
\multicolumn{4}{l}{Col. 2: Observation central frequencies}\\
\multicolumn{4}{l}{Col. 3: Robust parameter used for imaging}\\
 \multicolumn{4}{l}{Col. 4: UV taper used in Kilo-lambda.}\\
\multicolumn{4}{l}{$\ast$ Re-analysed 325 MHz data reported in \citet{venturi2013}}\\
\multicolumn{4}{l}{$\dagger$ Re-analysed 325 MHz data reported in \citet{Riseley2017A&A}}
\end{tabular}
\end{table}

We split the LL and RR correlations to obtain two data sets at two different frequencies i.e., 235 MHz and 610 MHz from the GMRT dual-frequency mode observations. For 325 MHz, combined RR-LL data was taken. Data analysis and imaging were done using the {\sc{SPAM}} pipeline (for details see \citealt{Intema_2017A&A}) using the parameters given in Table~\ref{tab:data-anal}. 'SPAM' is a powerful data analysis and imaging pipeline that takes care of direction-dependent variations (i.e., due to antenna beam pattern and due to ionosphere) in, visibility amplitude and phase, across the field of view. The initial flux and bandpass calibrations were done using the source 3C48 for the target sources Abell 2744, MACS0152 and MACS0025 and the source 3C286 for target source MACS1931. All the post-pipeline analysis were performed and measurements were taken using the Common Astronomy Software Applications (CASA) package \citep{McMullin_2007ASPC}.

High-resolution images were first made with robust $-1.0$ to map the point and bright sources and to check for any indication of diffuse radio emission in the imaged fields. To better image the extended emission, a Gaussian uv-taper was applied during imaging at both frequencies. Different robust parameters (see Table~\ref{tab:data-anal}) were used to obtain better diffuse emission or rms values as per the requirements of our study. As the observation of the object MACS0025 was spread over two days, two data sets were combined to make a single data set for imaging. The 325 MHz image of Abell 2744 was made with robust $-1$, and the resolution was lowered (convolved) to $35^{\prime\prime}\times35^{\prime\prime}$ to match the resolution of \citet{venturi2013}. Other archival data of cluster MACS0025 at GMRT 325 MHz was analysed with robust $0.5$, matching the parameters used for 235 and 610 MHz data for the same object. Data from only one session from \citet{Riseley2017A&A} was imaged in this study to keep the total observation time similar to ours (see Table~\ref{tab:radioobs-list}).

\subsection{Estimation of flux error and spectral index}\label{err-SI-comp}

We computed the flux densities ($S$) of observed radio sources within 3$\sigma$ contours ($\sigma$ is noise rms of the images)  and flux density errors ($\sigma_S$) using the usual relation

\begin{equation}\label{eq:Flux_err}
\sigma_S = \sqrt{(0.1 S)^2 + N ({\sigma})^2}
\end{equation}

where N is the number of beams covered by the total diffuse emission and $0.1S$ (i.e., 10\% of $S$) was assumed as the possible error due to calibration uncertainties.

The spectral index maps were created after re-gridding the two images of different frequencies using the IMREGRID task of CASA. The images used for making spectral index maps were made with uniform weights from data with matched uv-ranges of the different frequencies. Thereafter, we matched the resolution by convolving the beam of 610 MHz images to the beam of 235 MHz using the IMSMOOTH task. The images were then masked below $3\sigma$ using the 235 MHz image. Finally, we computed the spectral index using IMMATH with the relation 

\begin{equation}\label{eq:Sp-Ind}
\alpha = \frac{\log(S_{\nu_2}/S_{\nu_1})}{\log(\nu_2/\nu_1)}
\end{equation}

where $S_{\nu_\#}$ and $\nu_\#$ (with $\nu_1 > \nu_2$ ) are the flux densities and the respective observed frequencies.

Further, we calculated the spectral index error with IMMATH task using the relation given below \citep{Kim_2014JKAS}

\begin{equation}\label{eq:SP-error}
\alpha_{err}(\alpha_{\nu_2,\nu_1})=\frac{1}{\log(\nu_2/\nu_1)}\times\left[\frac{\sigma_{\nu_1}^2}{I_{\nu_1}^2}+\frac{\sigma_{\nu_2}^2}{I_{\nu_2}^2}\right]^{\frac{1}{2}}
\end{equation}

with $I$ as the total intensity at respective frequencies at each of the pixels. 

\section{Results}\label{result}

In Table~\ref{tab:results} we summarize our findings. We mention the detected radio structures at different frequencies in this study with respective rms ($\sigma$) value and beam sizes. We also report the angular sizes of the observed structures, corresponding flux densities with errors (using Eq.~\ref{eq:Flux_err}) and luminosities. For halos, the sizes are reported along North-South (NS) and East-West (EW), relic sizes are taken as Largest Linear Size (LLS) and the average width as given in the Table~\ref{tab:results}. Angular size, LLS and flux densities are estimated at $3\sigma$ contours of the respective sources. We also include here the  values of various parameters measured from the re-analysed GMRT 325 MHz archival data of two clusters, namely Abell 2744 and MACS0025. The results from the individual clusters are presented below.

\begin{table*} 
\caption{Radio properties of the sources} %
\label{tab:results}      
\centering          
\begin{tabular}{llccccrr}     
\hline\hline     
Cluster name & Source/ & Frequency & Beam size & rms & Size &   \multicolumn{1}{c}S  & Luminosity \\
& Emission& (MHz) && ($\mu$Jy beam$^{-1}$)  & & \multicolumn{1}{c}{(mJy)}&  \multicolumn{1}{c}{($10^{24}$ W Hz$^{-1}$)}\\
(1)&(2)&(3)&(4)&(5)&(6)&(7)&(8)\\
\hline  
Abell 2744  &  Halo &  610 & $20.62^{\prime\prime} \times15.83^{\prime\prime}$ PA $4.27^{\circ}$ & 100 & $300^{\prime\prime} \times 275^{\prime\prime}$ & $103.2\pm10.4$  & $31.8\pm3.2$ \\
& Relic~A &  &  & & $325^{\prime\prime} \times105^{\prime\prime}$ & $46.8\pm4.7$ & $14.4\pm1.5$ \\
& Relic~B & &   &&$190^{\prime\prime} \times35^{\prime\prime}$& $4.3\pm0.5$ & $1.3\pm0.2$ \\
&${\dagger}$\rdelim\{{3}{5pt} Halo &  325  & $35.0^{\prime\prime} \times35.0^{\prime\prime}$ PA $0.0^{\circ}$ & 450 & $370^{\prime\prime} \times 350^{\prime\prime}$ & $307.7\pm31.0$ & $90.0\pm9.0$\\
& $\;\;\;\;\;$Relic~A & & && $400^{\prime\prime} \times155^{\prime\prime}$& $118.7\pm12.1$ &  $35.6\pm3.6$ \\
& $\;\;\;\;\;$Relic~B & &  && $162^{\prime\prime} \times39^{\prime\prime}$& $8.6\pm1.3$ & $2.2\pm0.3$\\
&  Halo &  235  & $25.55^{\prime\prime} \times17.42^{\prime\prime}$ PA $-2.15^{\circ}$ & 360 & $275^{\prime\prime} \times 270^{\prime\prime}$ &$302.3\pm30.5$ & $93.2\pm9.4$   \\
& Relic~A & & && $325^{\prime\prime} \times100^{\prime\prime}$ & $123.0\pm12.6$ & $37.9\pm3.9$ \\
& Relic~B & &  && $215^{\prime\prime} \times45^{\prime\prime}$ & $12.6\pm1.8$ & $3.9\pm0.6$\\
\hline
MACS0152  & Relic ? & 610 & $18.89^{\prime\prime} \times15.91^{\prime\prime}$ PA $-2.43^{\circ}$  & 70 & -- & -- & -- \\
& Relic ? & 235 & $22.28^{\prime\prime} \times16.39^{\prime\prime}$ PA $-5.96^{\circ}$ & 500 &  $65^{\prime\prime} \times30^{\prime\prime}$ & $9.2\pm1.3$ &  $5.7\pm0.8$\\
\hline
MACS0025  &  Relic &  610 & $6.34^{\prime\prime} \times5.84^{\prime\prime}$ PA $34.66^{\circ}$ & 90 & $53^{\prime\prime} \times 9^{\prime\prime}$  & $2.0\pm0.3$& $2.8\pm0.4$\\
&  Relic &  325 & $13.52^{\prime\prime} \times10.91^{\prime\prime}$ PA $-65.41^{\circ}$ & 200 & $68^{\prime\prime} \times 26^{\prime\prime}$  & $5.1\pm0.8$& $7.2\pm1.1$\\
&  ?  &  235 & $15.53^{\prime\prime} \times12.17^{\prime\prime}$ PA $0.01^{\circ}$ & 700 & -- & -- & --\\
\hline
MACS1931  & Mini-halo &  610 &  $9.07^{\prime\prime} \times7.56^{\prime\prime}$ PA $5.83^{\circ}$ & 95 & $48^{\prime\prime} \times 54^{\prime\prime}$& $143.6\pm14.4$  & $60.5\pm6.0$\\
  & Mini-halo &  235 &  $17.46^{\prime\prime} \times12.19^{\prime\prime}$ PA $3.29^{\circ}$ & 1400 & $53^{\prime\prime} \times 62^{\prime\prime}$ & $650.9\pm65.2$ &$274.2\pm27.5$ \\
\hline
\hline
\multicolumn{8}{l}{{\bf Col. 1:} cluster name {\bf Col. 2:} Possible type of source or emission structures {\bf Col. 3:} Frequency of observation}\\ 
\multicolumn{8}{l}{{\bf Col. 4:} Image beam size in arc second. {\bf Col. 5:} Image rms ($\sigma$) in $\mu$Jy beam$^{-1}$ {\bf Col. 6:} Size of the structures}\\
\multicolumn{8}{l}{{\bf Col. 7:} Flux density in mJy {\bf Col. 8:} Luminosity of the sources in $10^{24}$ W Hz$^{-1}$}\\
\multicolumn{8}{l}{$\dagger$Size, flux density and Luminosities are calculated within 1 mJy contour to compare with \citet{venturi2013} values.} \\
\multicolumn{4}{l}{ (? means-proposed or doubtful)}\\
\hline
\end{tabular}\label{cluster-list}
\end{table*}

\subsection{Cluster MACS0014 or Abell 2744}

\begin{figure*}
\includegraphics[width=9cm]{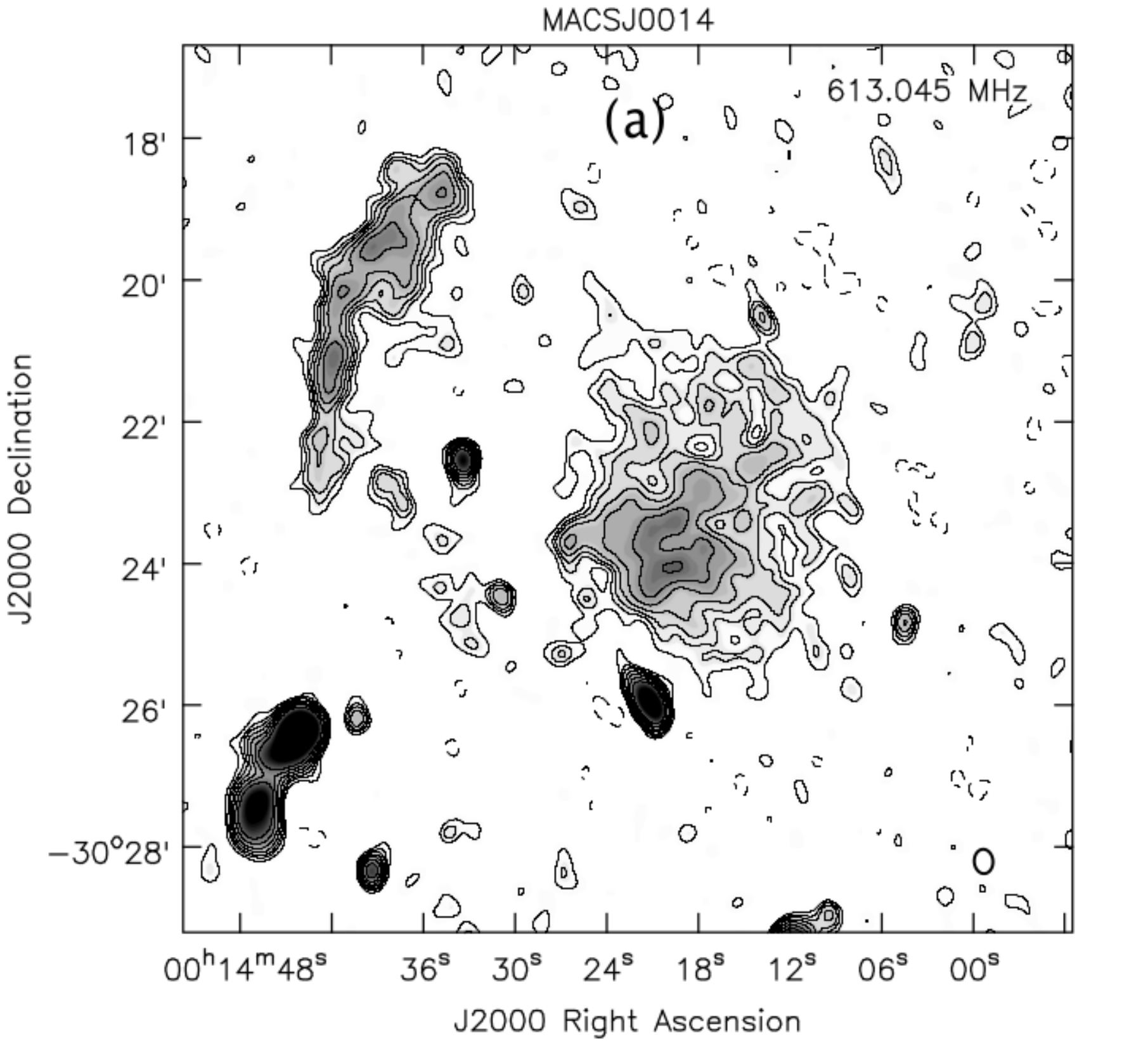}\hspace{-0.4cm}
 \includegraphics[width=9cm]{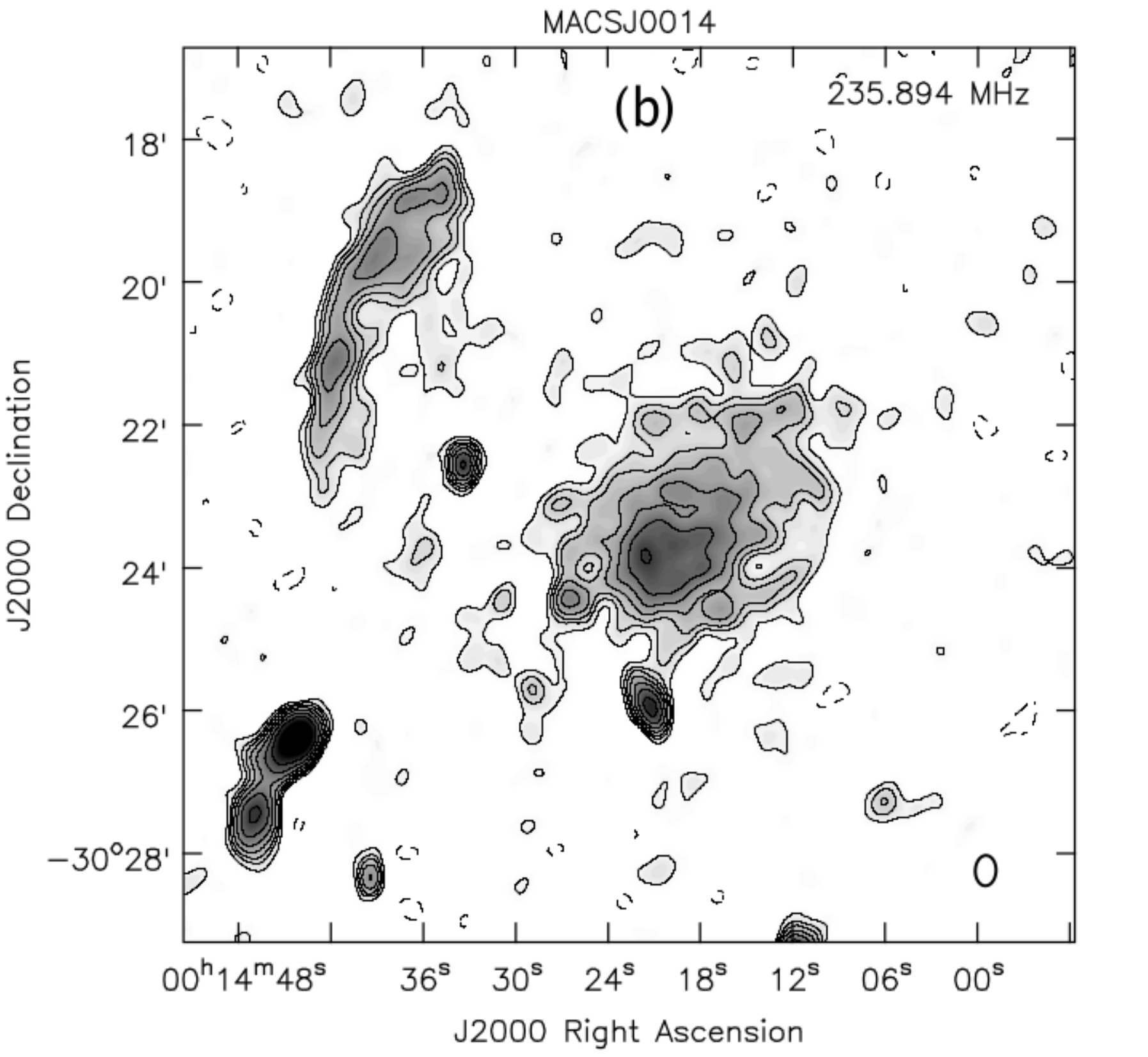}
\caption{ {\bf Panel~$(a)$:} 610 MHz GMRT radio continuum map of Abell 2744 plotted as gray colour and black contours are at the level of $3\sigma$ ($\sigma = 100 \;\mu$Jy~beam$^{-1}$) and further 6 contours are in multiplication of $\sqrt{2}$ of first contour. A negative contour at $-3\sigma$ has been plotted as dashed line. {\bf Panel~$(b)$:} Similar image and contours of the same cluster at GMRT 235 MHz. Here, $\sigma=360\;\mu$Jy~beam$^{-1}$.}\label{radio}
\end{figure*}

Abell 2744 is a moderately distant cluster at redshift $z=0.308$. This is a hot cluster with average temperature of $8.53\pm0.37$~keV \citep{Mantz_2010MNRAS}. Radius and mass are reported as $r_{500}= 1.65\pm0.07$~Mpc and $M_{500}=17.6\pm2.3\times10^{14}\;\rm{M_{\odot}}$ respectively \citep{Ebeling2010}. Importantly, the authors found no cooling flow in the cluster but, presence of substructures in X-rays indicate an ongoing merging process. This object is also well-known for hosting a radio halo and relics, as reported at high frequency with wideband VLA 1-4 GHz and 1.4 GHz \citep{Pearce_2017ApJ,Govoni2001A&A,Govoni2001}. At low frequency, 325 MHz at VLA \& GMRT \citep{Orr2007A&A,venturi2013} and from surveys \citep{George_2017MNRAS} at 88-200 MHz (The Galactic and Extra-Galactic All-sky MWA Survey (GLEAM), \citealt{Wayth_2015PASA}) as well as 150 MHz (TIFR-GMRT Sky Survey: Alternative Data Release (TGSS-ADR1), \citealt{Intema_2017A&A}).

We detect both the relics and an extremely large halo (Both Panels in Fig.~\ref{radio}) at GMRT 235 and 610 MHz. At 610 MHz, the image has rms ($1\sigma$) values of $100~\mu$Jy~beam$^{-1}$ with a beam of $20.62^{\prime\prime}\times 15.83^{\prime\prime}$ and PA~$4.27^\circ$. Similarly, at 235 MHz, the rms is 360~$\mu$Jy~beam$^{-1}$ with a beam of $25.55^{\prime\prime}\times 17.42^{\prime\prime}$ and PA~$-2.15^\circ$. The lowest contour is plotted at $\pm3\sigma$ and further contours are plotted at intervals of $\sqrt{2}$ of the first positive contour (see Fig.~\ref{radio}).

\subsubsection{Radio halo}\label{halo-obs}

\begin{figure}
\includegraphics[width=8.6cm]{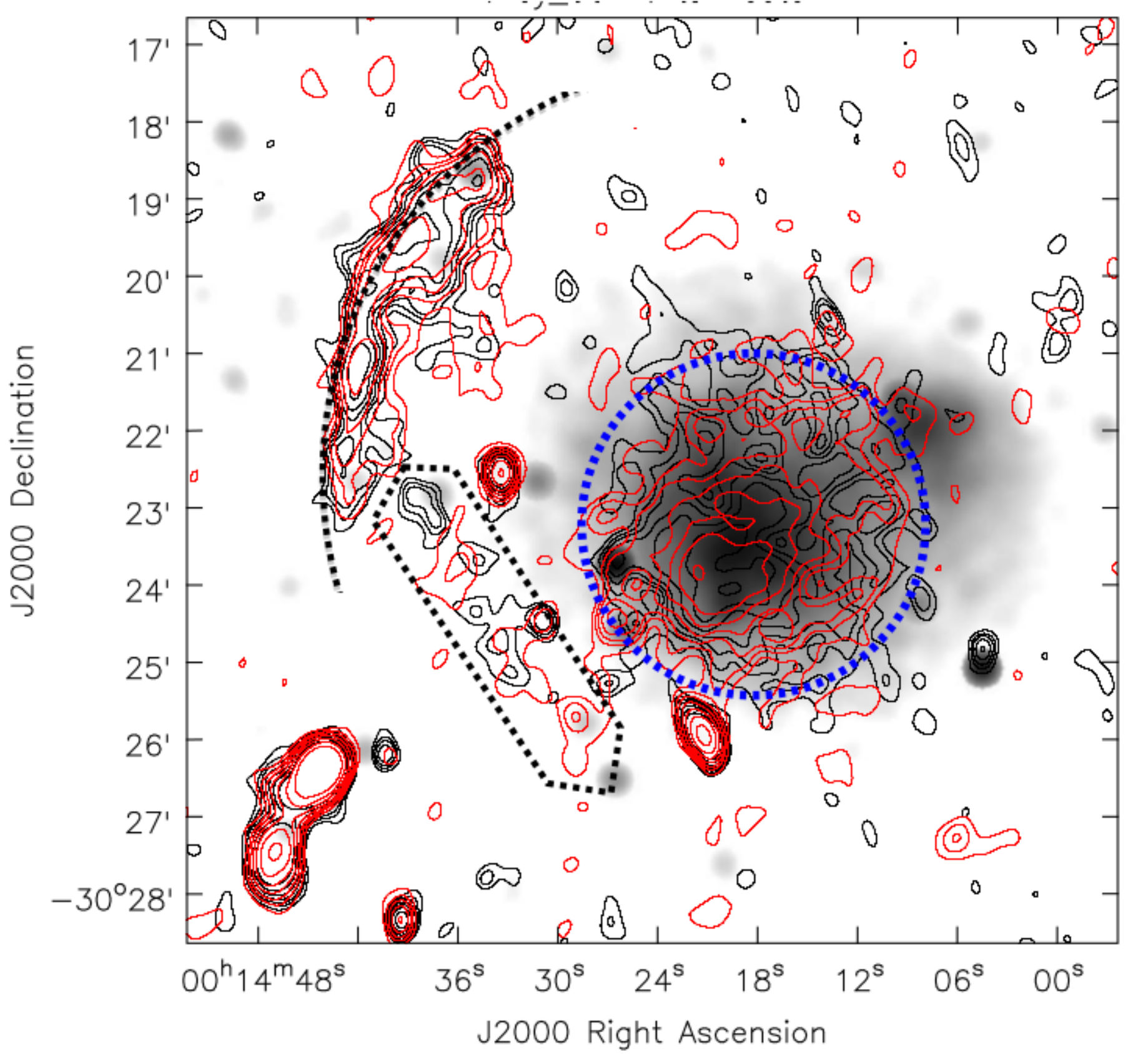} 
\caption{Radio contours for Abell 2744 are plotted together for 610 (black) and 235~MHz (red) with the same levels as Figure~\ref{radio}. Blue dashed circle encloses the radio halo, black dashed region shows the relic~B and almost concentric circular arc with dashed black line indicates the outer edge of the relic~A. X-ray photon count is plotted as gray colour in the background.}\label{A2744_comb}
\end{figure}

The radio halo in this cluster is a textbook halo (see \citealt{Weeren_2019SSRv} for halo definition), truly representing its class having roughly a regular morphology with almost circular shape (see Fig.~\ref{A2744_comb}, blue dashed circle). The halo is almost similar in shape and morphology at both frequencies, as it can be noticed in the overlapped map in Figure~\ref{A2744_comb}. However, the sub-structures at both frequencies differ in shape and morphology. At 610~MHz, a split in the halo centre can also be found (see Fig.~\ref{radio}(a)). The radio halo peak is slightly ahead of X-ray peak towards SE, but the total extension are almost the same (see Fig.~\ref{A2744_comb}). This radio halo has a projected angular size (LLS i.e. the Largest Linear Size) of about 6$\arcmin$ ($\sim$~1.6 Mpc) as found at $3\sigma$ detection limit at both frequencies (See Table~\ref{tab:results}). Our re-analysed GMRT 325~MHz map measures a larger LLS of about 1.9~Mpc (image not shown), consistent with the reported values of $\sim 1.9$~Mpc \citet{venturi2013}, both measured within the contour level of 1~mJy~beam$^{-1}$. The flux densities at 235 and 610 MHz respectively are given by $302\pm30$ and $103\pm10$~mJy. At 325~MHz, our measured value is $308\pm31$~mJy (Detail in Table~\ref{tab:results}), slightly lower than the flux density of $323\pm26$~mJy reported in \citet{venturi2013}.

\subsubsection{Radio relic}\label{relAB-obs}

The most prominent radio relic (Relic~A as shown in Fig.~\ref{A2744_comb}) in this cluster is a peripheral, single, bow-shaped relic with concave side facing towards the cluster centre. It follows the curvature of the radio halo structure, as shown in Fig.~\ref{A2744_comb} with almost concentric circles (dashed Black and Blue). The relic~A has several sub-structures and is not sharp edged like the CIZA J2242.8+5301 cluster \citep{Stroe_2013A&A}. It has almost the same length as the halo 5.5$\arcmin$ ($\sim$~1.5~Mpc) at 235 and 610 MHz but 1.8 Mpc at 325 MHz with a large average width of about 500 kpc in all the reported frequencies (see Table~\ref{tab:results}). The relic is placed at about 1.7~Mpc ($365^{\prime\prime}$) away from the cluster centre i.e., at about $r_{500}$ radius of the said cluster. This relic is a bright radio source with flux densities $46.8\pm4.7$, $118.7\pm12.1$ and $123.0\pm12.6$~mJy  at 610, 325 and 235~MHz respectively.

A second relic, Relic~B, as shown in Figure~\ref{radio}, is detected just in front of the bigger merging group, moving towards the South-East (SE) corner (in RA-DEC plane). This relic is faint and broken into parts (as seen above $3\sigma$ level). Broken parts in different frequencies do not coincide, rather complement each other to make a full relic structure when over-plotted, as shown in Figure~\ref{A2744_comb}, indicated as Relic~B within the black dashed area. \citet{Pearce_2017ApJ} could detect the full relic only when imaged with wide band VLA~1-4~GHz.\citet{venturi2013} have also detected a part of it in their GMRT 325 MHz deep images, and identified it as a radio bridge between halo and the Relic~A. Since, the Relic~B is not a continuous structure (at the $3\sigma$ level) at any frequency, we measured the flux from a common polygonal area. The respective flux densities are $5.8\pm1.0$, $12.1\pm2.1$ and $24.8\pm3.7$~mJy i.e., fainter by almost an order in magnitude compared to the Relic~A.

\subsubsection{Spectral index}\label{Spec-ind1}

\begin{figure*}
\includegraphics[width=9cm]{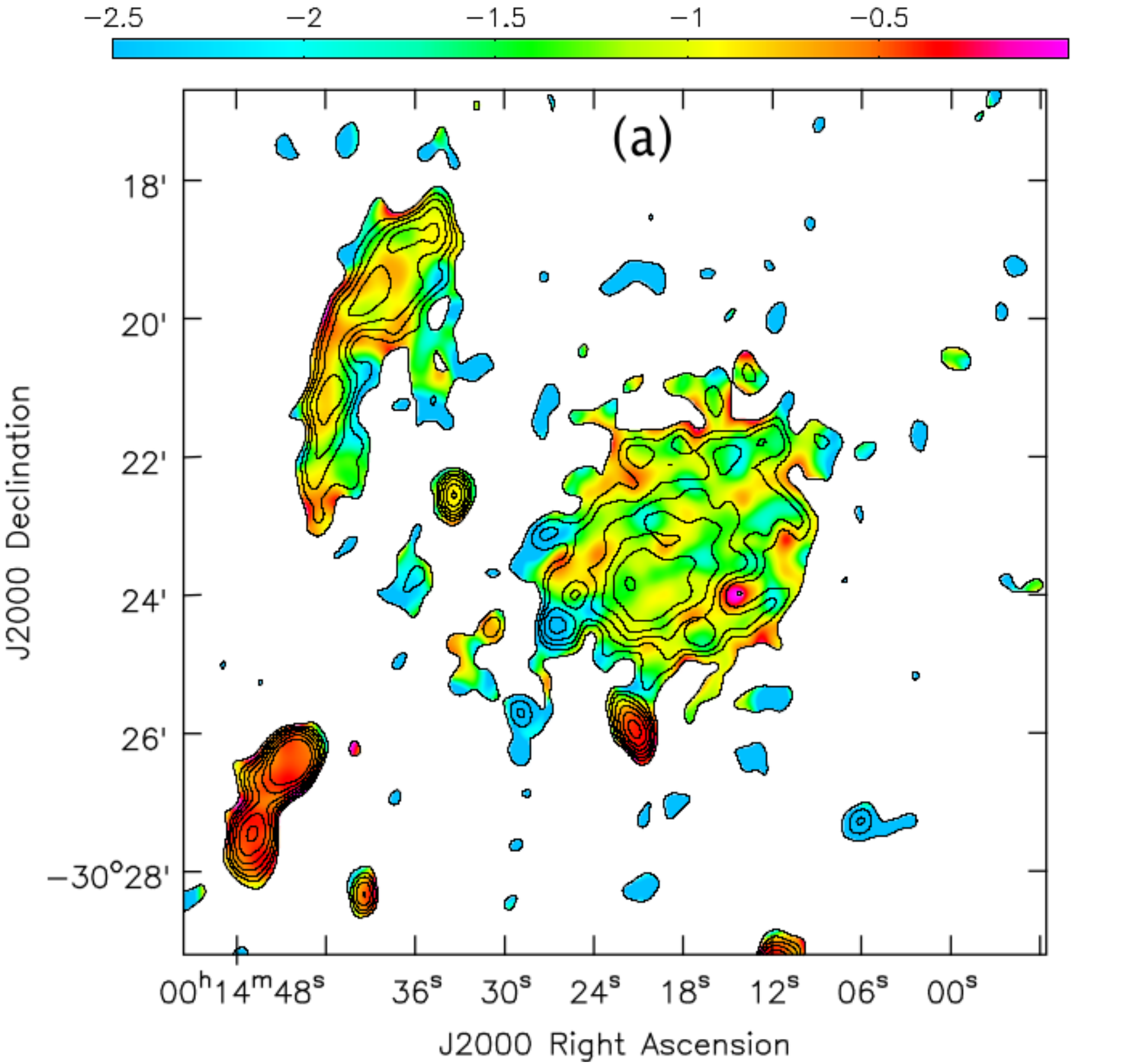}\hspace{-0.4cm}
 \includegraphics[width=9cm]{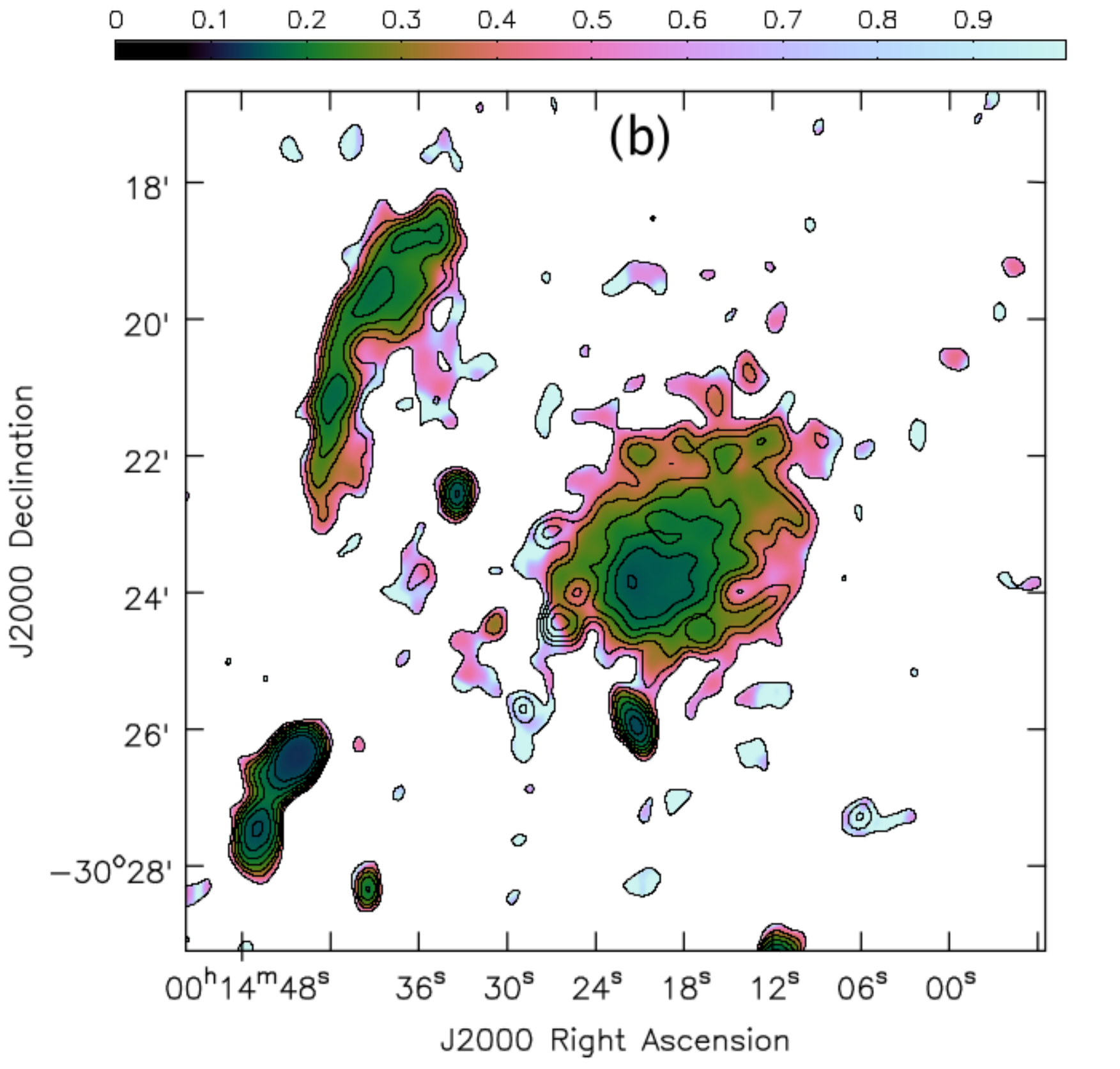}
\caption{ {\bf Panel~$(a)$:} Spectral index (see Eq.~\ref{eq:Sp-Ind}) map of Abell 2744 in colour scale. {\bf Panel~$(b)$:} Spectral index error (see Eq.~\ref{eq:SP-error}) map with colour scale. Radio flux contours in both panel are same as Fig.~\ref{radio}.}\label{radio-spec-1}
\end{figure*}

\begin{table} 
\caption{Total spectral index of the detected structures} %
\label{tab:spectra}      
\centering          
\begin{tabular}{|llcc|}     

\hline
\multicolumn{4}{c}{(235-610 MHz)}\\
\hline       
Cluster name & Halo/relic & Spectral index & Error   \\
\hline
Abell 2744  & Halo & -1.17 &  $\pm0.33$\\
 & Relic~A & -1.14 & $\pm0.36$\\
 & Relic~B & -1.65 & $\pm0.62$\\
\hline
\multicolumn{4}{c}{(235, 325 \& 610 MHz)}\\
\hline
Abell 2744  & Halo & -1.21 &  $\pm0.20$\\
 & Relic~A & -1.09 & $\pm0.20$\\
 & Relic~B & -1.34 & $\pm0.16$\\
\hline
\multicolumn{4}{l}{{\bf Col. 1:} Cluster name {\bf Col. 2:} Type of radio structures}\\ 
\multicolumn{4}{l}{{\bf Col. 3:} Average spectral index of the total structures}\\
\multicolumn{4}{l}{{\bf Col. 4:} Error in spectral index}
\end{tabular}\label{spectral-ind}
\end{table}

A spectral index map of 610 and 235~MHz emission is produced using the method described in Section~\ref{err-SI-comp} and shown in Figure~\ref{radio-spec-1}$(a)$. The corresponding error map computed using Eq.~\ref{eq:SP-error} is shown in Figure~\ref{radio-spec-1}$(b)$. It can be observed that the radio halo has a relatively flat spectrum with average value of about $\alpha_{halo}{^{610}_{235}} = -1.17\pm0.33$. Except few small scattered regions, mostly the spectrum is uniform ($\sigma\sim 1.1$) across the halo. The average or integrated spectral index of the relic~A is found to be $\alpha_{relA_{int}}{^{610}_{235}}= -1.14\pm0.36$. On the outer edge the relic has a very flat spectrum of $\alpha_{rel(A-E)}{^{610}_{235}}=-0.69\pm0.21$. This steepens gradually towards the inner parts of the relic to about  $\alpha_{rel(A-O)}{^{610}_{235}}=-1.29$. Therefore, it follows the usual continuous injection model reasonably i.e., $\alpha_{inj}{^{610}_{235}}=(\alpha_{int}{^{610}_{235}}-0.5)=0.64\sim \alpha_{rel(A-E)}{^{610}_{235}}$. Relic~B is on the other hand a very steep spectrum relic with average spectral index of $\alpha_{relB}{^{610}_{235}}=-1.65\pm0.62$.

\subsection{Cluster MACS0152}\label{macs0152-des}

\begin{figure*}
\includegraphics[width=8.8cm]{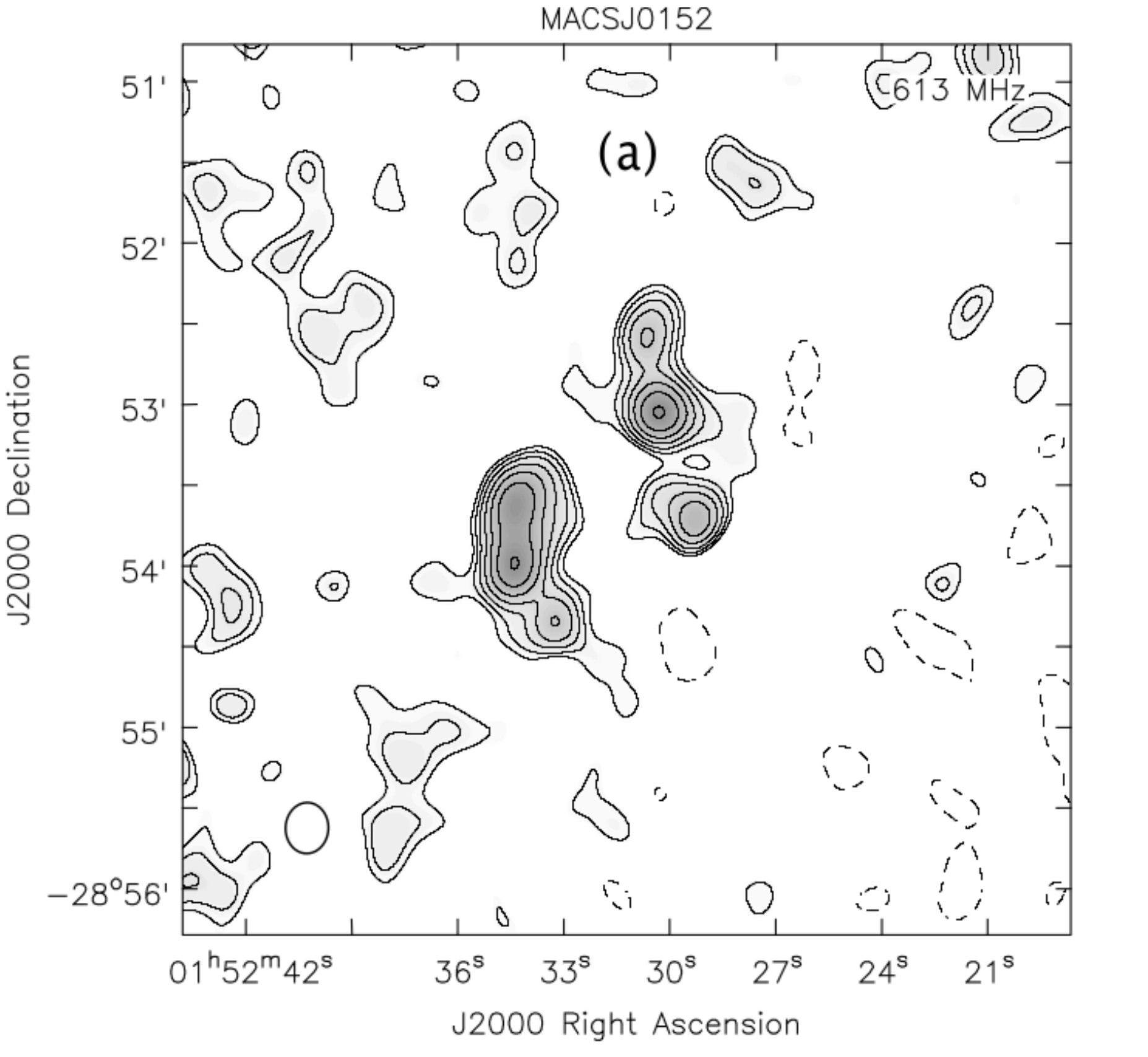}
\hspace{-0.4cm}
\includegraphics[width=8.8cm]{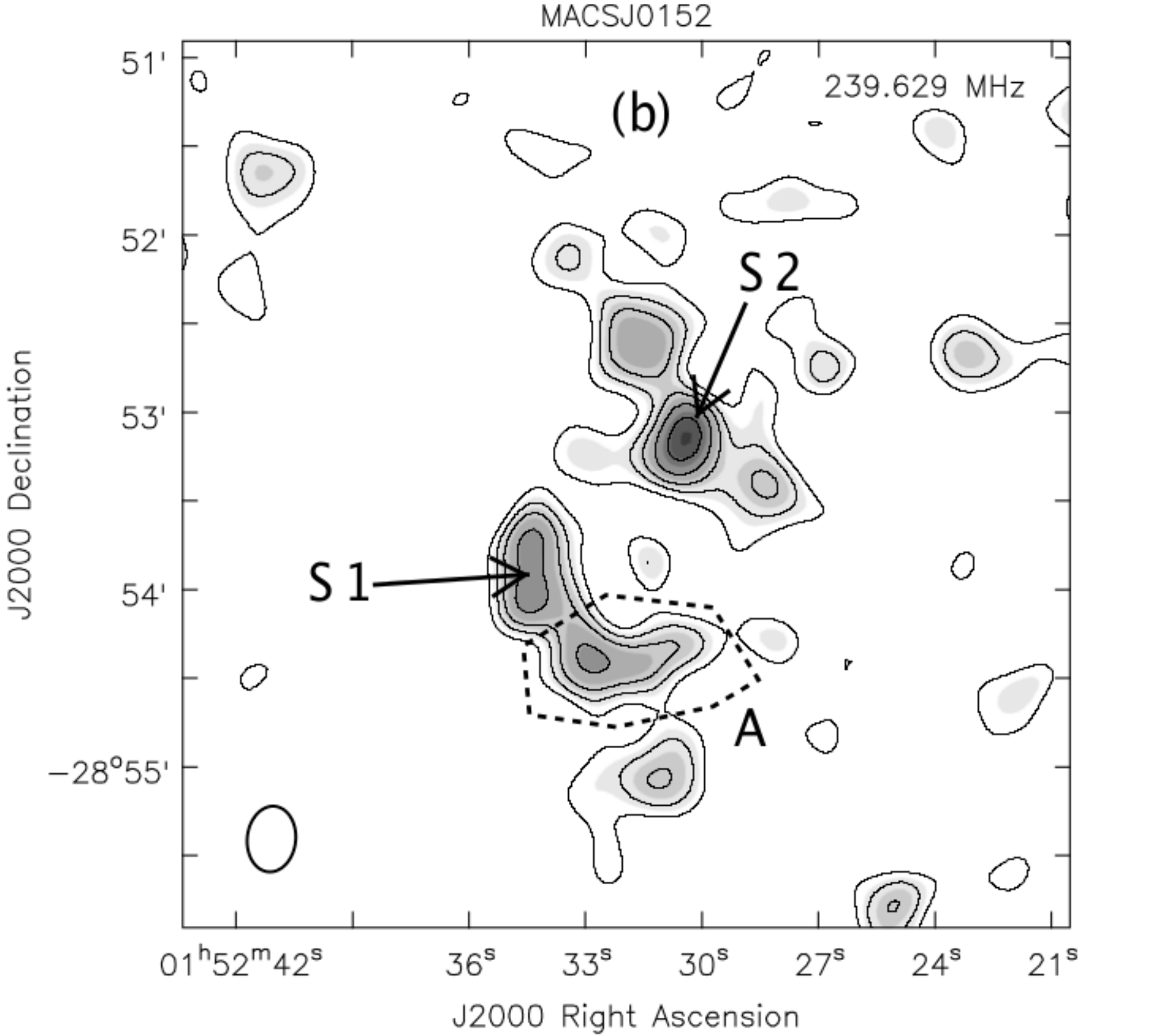}
\caption{{\bf Panel~$(a)$:} \& {\bf Panel~$(b)$:} Same contour levels as Figure~\ref{radio} for the cluster MACS0152 with rms values at 610~MHz is $\sigma=70\mu$Jy~beam$^{-1}$ and at 235 MHz it is $\sigma=500\mu$Jy~beam$^{-1}$. S1 and S2 are the two point sources having optical and X-ray counter part. Dashed line encloses the extended diffuse relic (?).}\label{mac0152}
\end{figure*}

\begin{figure*}
\includegraphics[width=8.8cm]{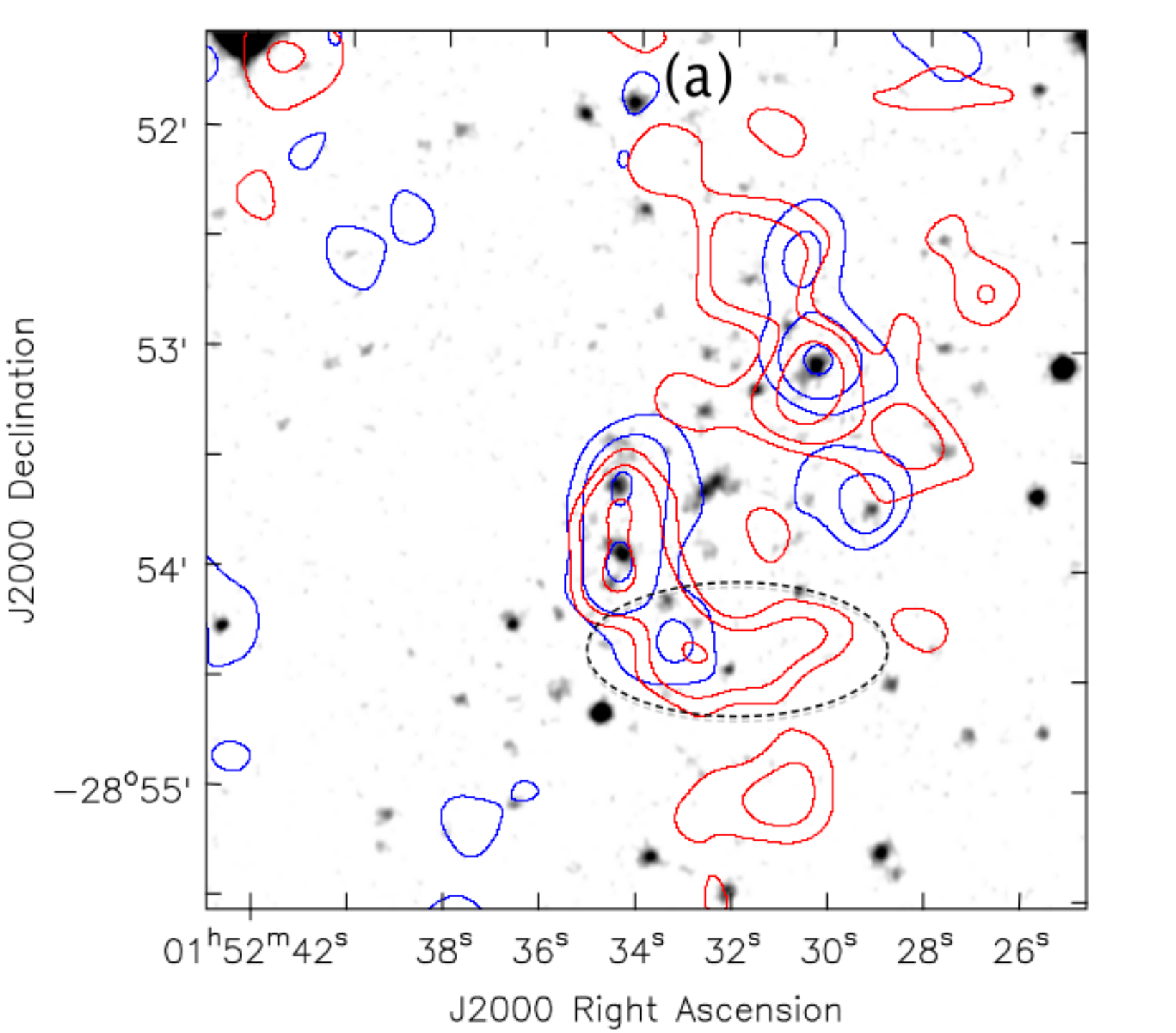} \hspace{-0.4cm}
\includegraphics[width=8.5cm]{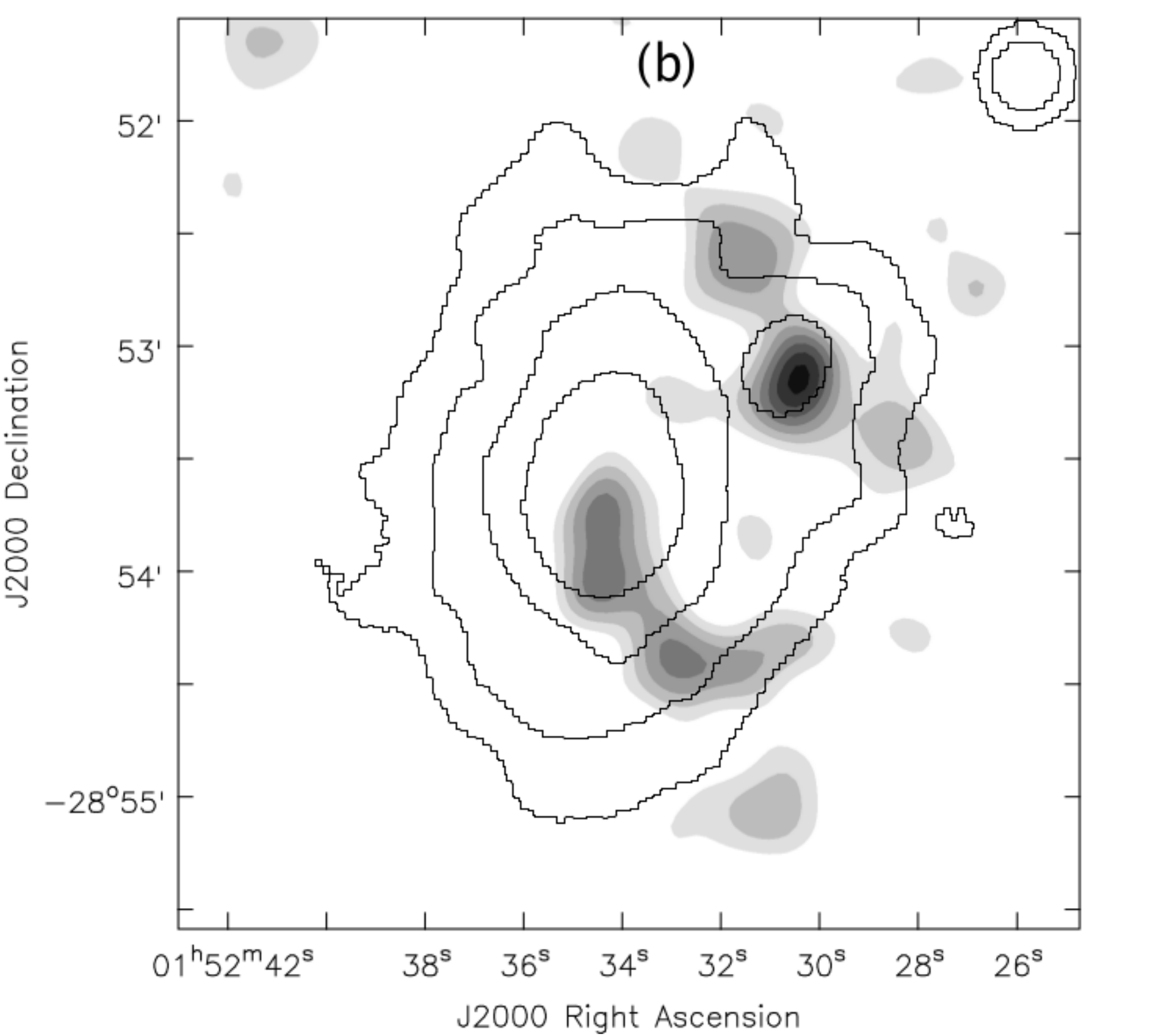}
\caption{{\bf Panel~$(a)$:} GMRT 610 (blue) and 235 MHz (red) contours (only 3 at 3,6,12 $\sigma=70\&500 \mu$Jy~beam$^{-1}$) are over plotted on SDSS optical image. {\bf Panel~$(b)$:} Chandra X-ray photon count map (0.5-7.0~keV, Gaussian smoothed) with black contours at 4,8,16 and 32 times $10^{-8}$~photons~s$^{-1}$~cm$^{-2}$~pixel$^{-1}$ and are over plotted on the gray colour radio map of cluster MACS0152 at GMRT 235~MHz.}\label{MACSJ0152-opt-xray}
\end{figure*}

MACS0152 is a relatively high redshift cluster at $z=0.413$, having  radius, $r_{500}=1.21\pm0.11$~Mpc, and mass at $M_{500}=7.9\pm2.2$ $10^{14}\;\rm{M_{\odot}}$ with a moderate temperature of $4.7\pm0.5$~keV. The cluster shows non-concentric X-ray contours towards the North-West (NW) direction with the signature of substructures or disturbed morphology indicating a probable merger \citep{Ebeling2010}. 
 
Our GMRT observation at 235~MHz has revealed a diffuse emission of about 0.5~Mpc size in this cluster. A one-sided arc-shaped radio emission is placed on the southern side of the cluster starting near to the centre to the half-way to the virial radius, marked as dashed area `A' in Figure~\ref{mac0152}$(b)$. The source could not be recovered fully in 610~MHz (Fig.~\ref{mac0152}$(a)$). Except a few unrelated small regions of radio emission at 235~MHz, no other prominent peripheral or arc/bow shaped relic or halo emission has been observed at both frequencies. Among the radio sources, only two radio bright spots marked as `S1' and `S2' in Figure~\ref{mac0152}$(b)$ are of point source origin and are correlated to the optical and bright X-ray counterparts (See Fig.~\ref{MACSJ0152-opt-xray}$(a)\&(b)$), they are the radio and X-ray bright galaxies. One (marked as `S1') is connecting to the left arc-like radio structure~`A'. Coincidentally, above mentioned two bright sources are placed at the centre of the two galaxy groups (X-ray sources: \citealt{Ebeling2010}) that are supposed to be colliding with each other. Remaining part (i.e., other than `S1' region) of the left radio structure has no optical or X-ray counterparts of point source origin and are purely diffuse in nature (see Fig.~\ref{MACSJ0152-opt-xray}$(a)\&(b)$). The total flux density of this arc-like structure at 235~MHz is  $9.2\pm1.3$~mJy (radio power of $5.7\pm0.8\times10^{24}$~W~Hz$^{-1}$) and spectral index of the part of the source that has been observed in both 610 and 235 MHz is extremely steep $\alpha_{rel}=-2.3\pm0.6$. Size of this diffuse radio structure is about 1.5$\arcmin$, that translate to a linear size of about 0.5~Mpc and the structure, as projected on the sky, is positioned well inside the virial radius of the cluster.

\subsection{Cluster MACS0025}\label{macs0025-des}

The Cluster MACS0025 is a distant cluster with redshift of $z=0.584$. This is a very hot and massive object having average temperature of $7.10\pm0.70$~keV as computed from the Chandra X-ray data \citep{Ebeling_2007ApJ}, and the total gravitating mass, $M_{500}=8.44\pm3.16\times10^{14}~\rm{M_{\odot}}$ \citep{Riseley2017A&A}. It has been reported from weak and strong lensing studies that two dark matter sub-groups with similar masses (about $2.5\times10^{14}~\rm{M_{\odot}}$) are separated from baryonic over-density regions, probably indicating a merger \citep{Brada2008ApJ} or a core oscillation phase \citep{Ma_2010MNRAS}. High X-ray brightness of $8.8\times10^{44}$~erg~s$^{-1}$ with presence of substructures also indicate dynamical activity in this cluster. 

\begin{figure*}
\includegraphics[width=8.5cm]{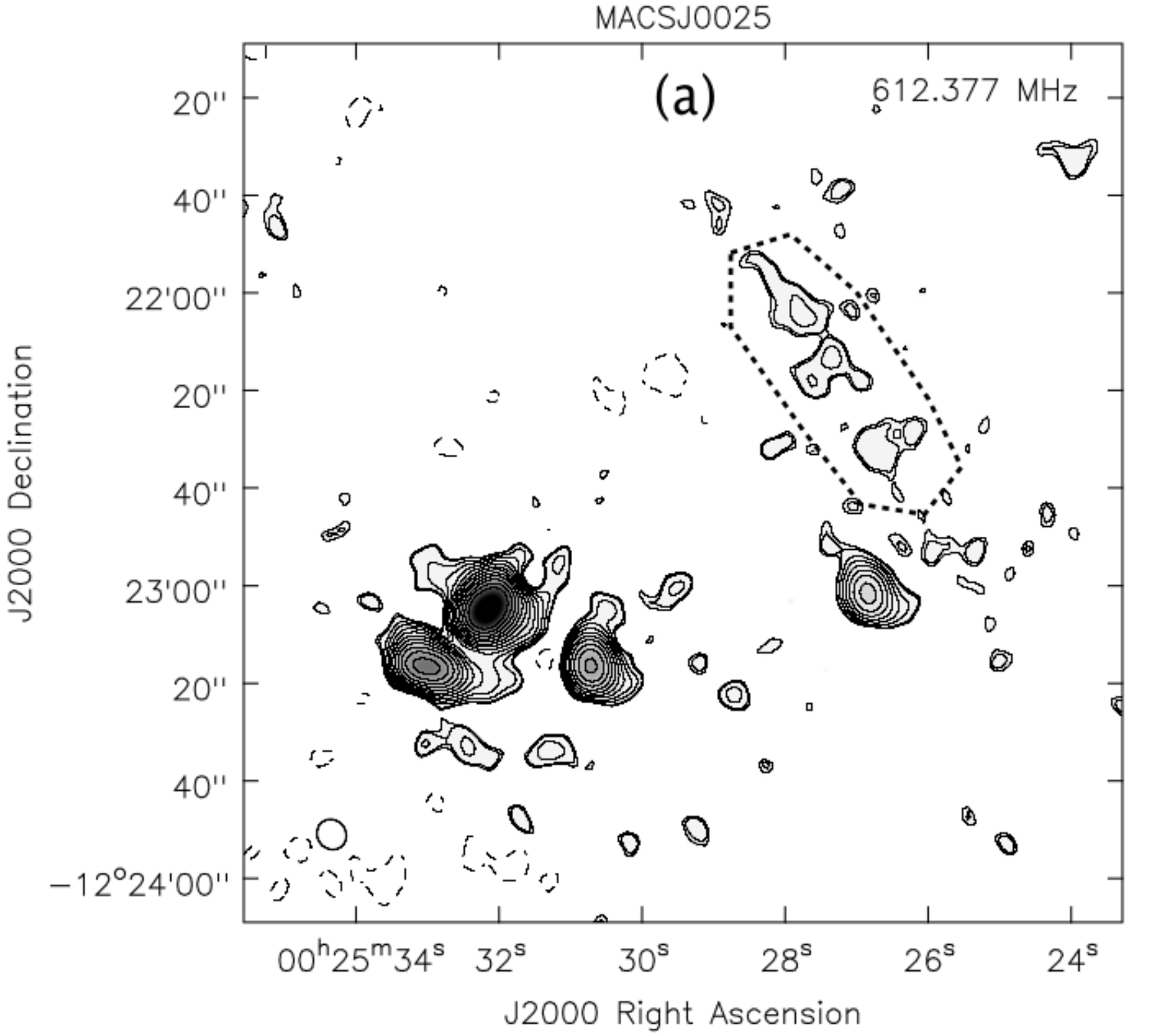}
\hspace{-0.4cm}
\includegraphics[width=8.8cm]{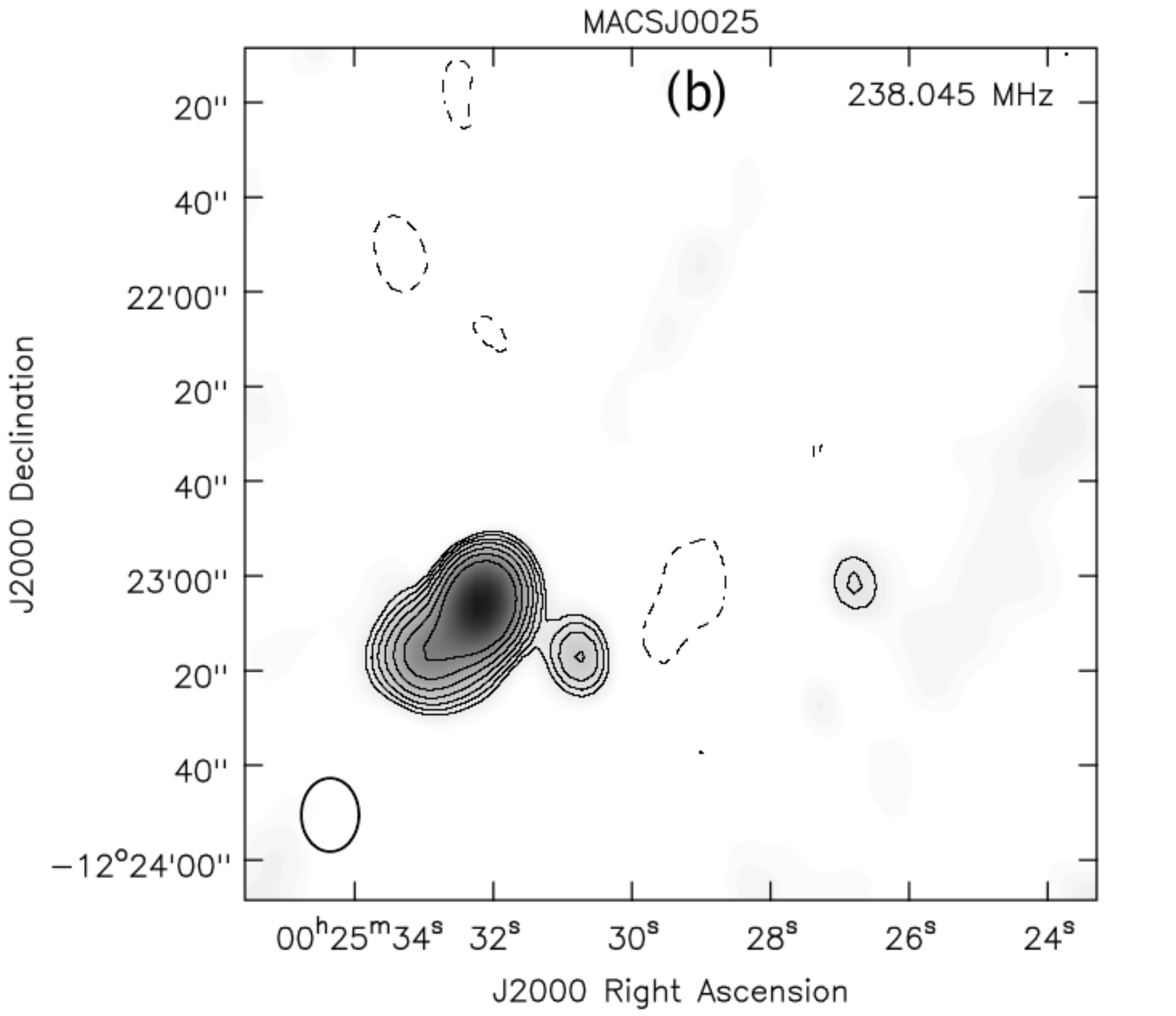}
\caption{{\bf Panel~$(a)$:} \& {\bf Panel~$(b)$:} 610 and 235~MHz map for the cluster MACS0025 with first contour at 2.5$\sigma$, rest of the contours and colour levels are same as Figure~\ref{radio}. The rms values at 610 MHz is $\sigma=90\mu$Jy~beam$^{-1}$ and at 235~MHz it is $\sigma=700\mu$Jy~beam$^{-1}$ with beam sizes as given in Table~\ref{tab:results}.}\label{mac0025}
\end{figure*}

\begin{figure}
\includegraphics[width=8.8cm]{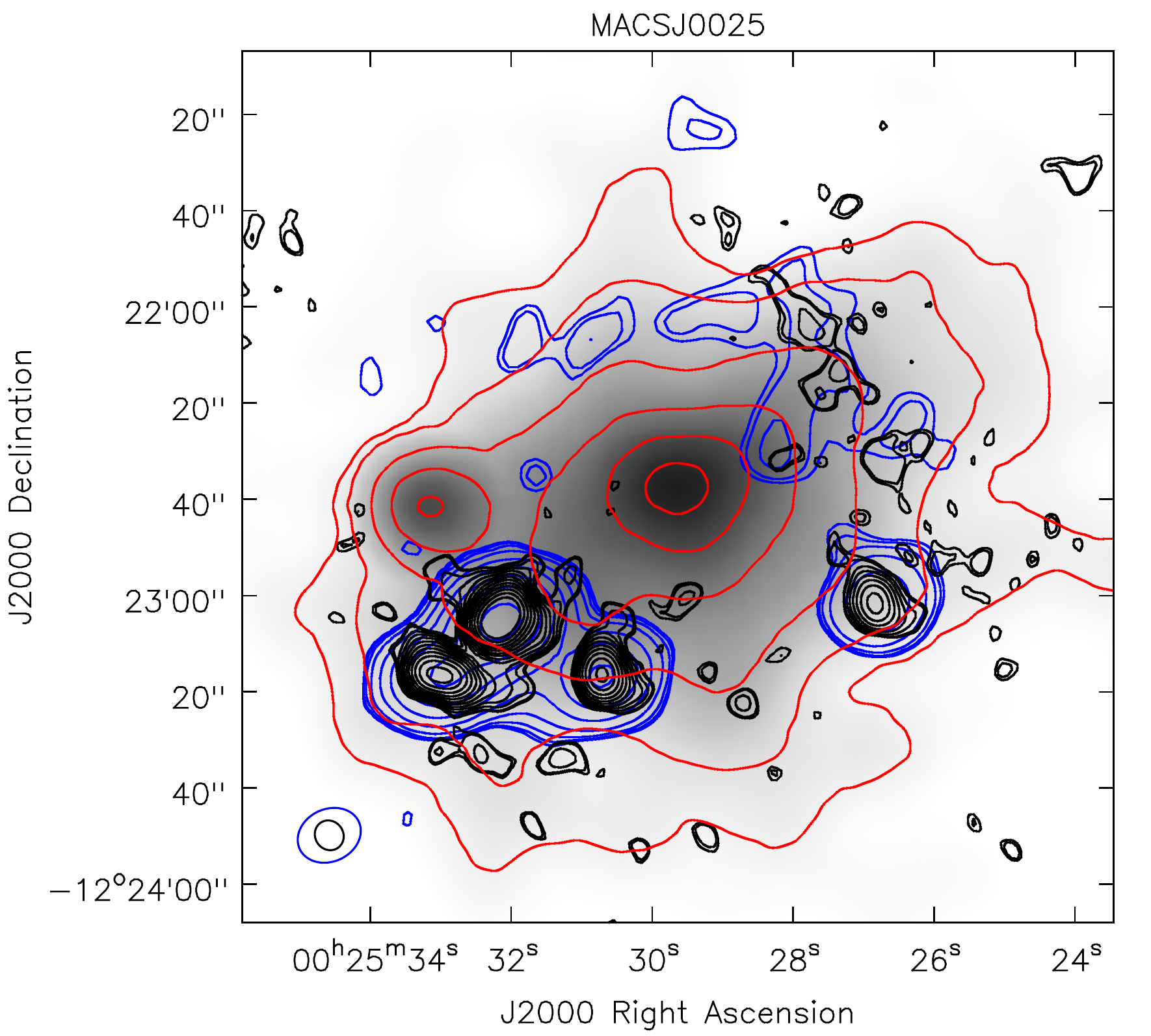}
\caption{Chandra X-ray photon count (0.5-7.0 keV, Gaussian smoothed) map of cluster MACS0025 has been plotted in gray scale and contour map is over-plotted in red colour with contours at 1.3, 2, 4, 6, 8 and 9.5 times $2.87\times10^{-8}$ photons s$^{-1}$ cm$^{-2}$ pixel$^{-1}$. Radio contours are in blue for 325~MHz map with controus at 3, 3.5, 4, 6, 12, 24, 48, 96 times the 1$\sigma$ rms (where $\sigma=200\mu$Jy~beam$^{-1}$) and black for 610 MHz map and contours are same as the Figure~\ref{mac0025}$(a)$.}\label{mac0025-xray}
\end{figure}

Evidence of diffuse radio emission, in particular relics in this cluster, was first reported by \citet{Riseley2017A&A} at 325 MHz using the GMRT. Our GMRT observation at 610 MHz has also revealed a very faint, relic like structure towards NW of the cluster as shown in the Figure~\ref{mac0025}$(a)$, marked with dashed line. The same structure cannot be seen in our 235~MHz map (Fig.~\ref{mac0025}$(b)$) having much higher rms caused by the data loss due to bandpass filter and presence of a few bright sources in the central field. Achieved final rms is just 700~$\mu$Jy~beam$^{-1}$ at 235 MHz, as compared to a much lower i.e., 90~$\mu$Jy~beam$^{-1}$ at 610~MHz using the data analysis parameters mentioned in Table~\ref{tab:data-anal}. The relic source is an extremely faint object at 610~MHz with flux density of $2.0\pm0.3$~mJy. Our re-analysed image at 325 MHz has also shown a relic like structure around the same region as in 610~MHz map (see Fig.~\ref{mac0025-xray}). Apart from this relic, only 4 bright galaxies associated with some diffuse emission can be found in both frequencies, no radio halo or any other prominent relic or diffuse emission is detected in our maps. 

Radio contours are drawn on the Chandra archival X-ray map in Figure~\ref{mac0025-xray}. It can be noticed that the diffuse radio emission detected at 610 and 325~MHz in this study falls at the outskirts of the X-ray core but well inside the actual extent of the cluster ($r_{500}=1.15$~Mpc). The distance between the relic and the cluster centre is measured to be $45^{\prime\prime}$ i.e., 300 kpc only. The detected relic is just $53^{\prime\prime}$ in angular size which at this high redshift of $z=0.584$ would translate to a linear size of 350~kpc. In the re-analysed 325 MHz maps, the relic measures as $68^{\prime\prime}$. Spectral index of $\alpha^{610}_{325} = -1.03\pm0.41$ is measured within a common region of the relic emission at 610 and 325~MHz.

\subsection{Cluster MACS1931}\label{MJ1931}

\begin{figure}
\includegraphics[width=8.5cm]{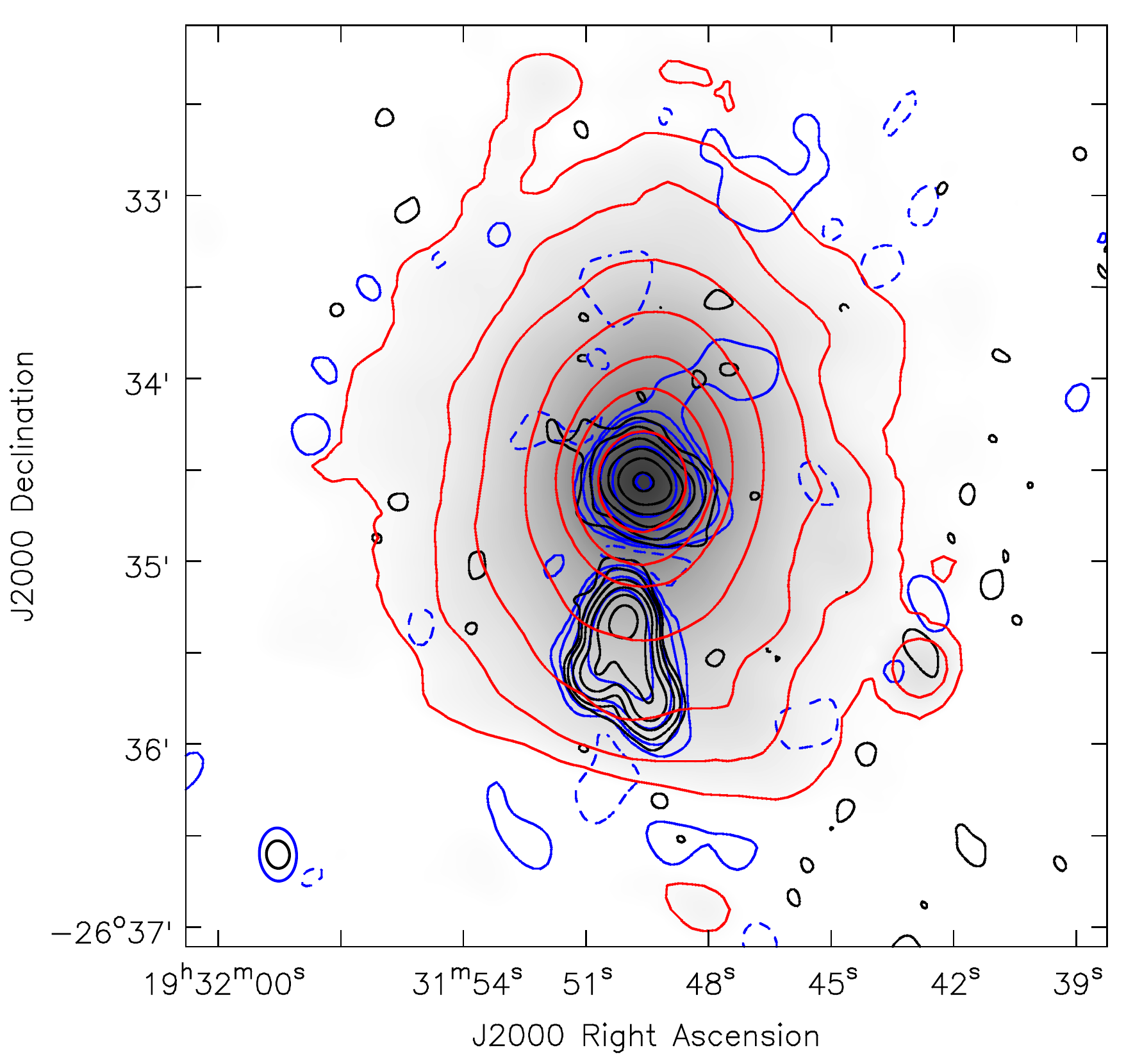} \hspace{-0.4cm}
\caption{610~and~235~MHz contour of the cluster MACS1931 plotted in black and blue respectively. Contours are drawn at the level of $\pm3\sigma$ (where rms i.e. $\sigma = 95\mu$Jy and $1400\mu$Jy~beam$^{-1}$ and negative contour in dashed line) and rest of the positive contours are at 9,27,81,243,729 $\sigma$. Red contours are the Chandra X-ray photon count map of 0.5-7.0~keV (Gaussian smoothed).}\label{Non-detect}
\end{figure}

Cluster MACS1931 at redshift z=0.352 is a very hot and massive cluster with temperature $T=8.36\pm0.39$ keV  \citep{Liu_2018MNRAS} and mass $M=9.94\times 10^{14}\rm{M_{\odot}}$ \citep{Ehlert_2013MNRAS}. This cluster host a BCG and having a highly X-ray bright ($\sim 1 \times 10^{45} \rm{erg\; s^{-1}}$) cool-core that recently got disturbed due to activity at the core \citep{Ehlert_2011MNRAS}. In radio waves, \citet{Giacintucci_2014ApJ} have mentioned it as a radio mini-halo candidate (uncertain), inferred from their 1.4 GHz VLA observations.

\begin{figure*}
\includegraphics[width=8.7cm]{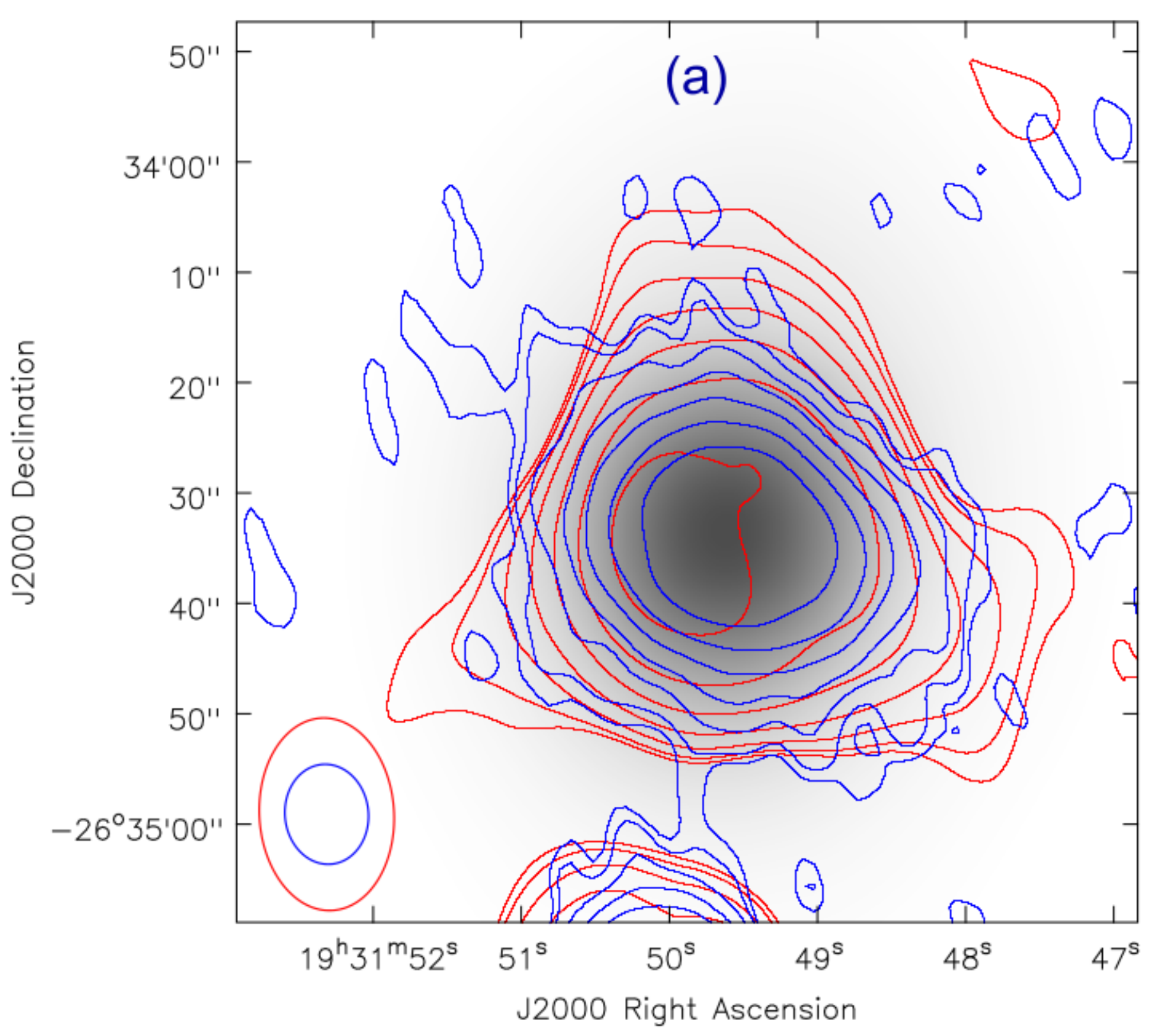}
\includegraphics[width=6.0cm]{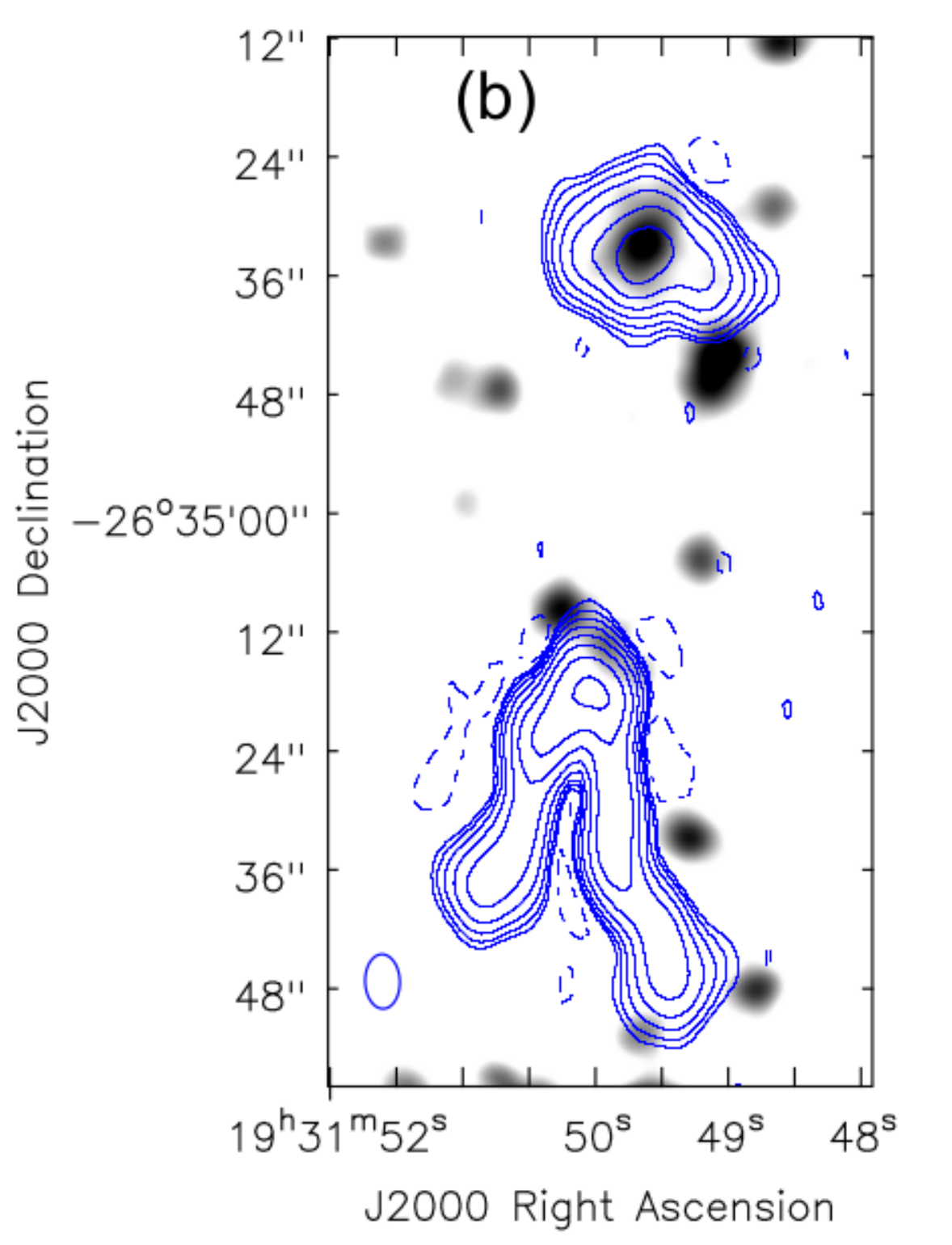} 
\caption{{\bf Panel~$(a)$:} 235 (red) and 610~MHz (blue) contour map of point source removed image of the cluster MACS1931 with contours at 5, 20, 80, 320 $\sigma$. Beam size and rms ($\sigma$) are $17.46^{\prime\prime} \times12.19^{\prime\prime}$ PA $3.29^{\circ}$, 1.2 mJy and $9.07^{\prime\prime} \times7.56^{\prime\prime}$ PA $5.83^{\circ}$, 0.1 mJy at 235~and~610~Mhz respectively. {\bf Panel~$(b)$:} High resolution 610 MHz contours plotted over gray scale optical SDSS image. Contours are at $\pm3$,9,27,81,243,729 $\sigma$, with $\sigma=0.09$mJy) and beam = $5.50^{\prime\prime} \times3.50^{\prime\prime}$ PA $1.83^{\circ}$.}\label{MiniHalo-jet}
\end{figure*}

In this study, the maps in Figure~\ref{Non-detect} have rms of $95$ and $1400\;\mu$Jy~beam$^{-1}$ respectively at 610 and 235~MHz with image details as given in Table~\ref{tab:results}. Both at GMRT 235 and 610 MHz, the cluster hardly shows any large diffuse radio emission that would amount to a halo (See Fig.~\ref{Non-detect}). It is also unlikely, as no major merging activity has been detected in this cluster. Rather, Chandra X-ray photon counts map shows comparatively smooth gas distribution indicating a relaxed state. X-ray emission is mostly dominated by the BCG, perfectly coinciding with radio galaxy core (Fig.~\ref{Non-detect}). But, these maps, at both frequencies show an extended emission, much beyond the typical size of the BCGs (see Table~\ref{tab:results}). By removing the central bright source (BCG), we re-imaged the diffuse emission following the method described in \citet{Weeren_2014ApJ} and recovered a significant diffuse emission at both the frequencies extended far from the cluster centre upto around 300 kpc (see Fig.~\ref{MiniHalo-jet}). This is a very bright diffuse radio source with fluxes measured as $651\pm65$ mJy and $144\pm14$ mJy with BCG flux densities $890\pm11$ mJy and $41\pm2$ mJy at 235 and 610 MHz respectively. Spectral index estimated to be $\alpha^{235}_{610}=-1.55\pm0.21$. Radial radio brightness profiles observed to follow the X-ray brightness profile reasonably.

This cluster also shows an interesting radio galaxy with extremely bent jets with a projected jet separation angle of about $30^{\circ}$ as seen at 610~MHz (see Fig.~\ref{MiniHalo-jet}(b)), a typical Narrow Angle Tail (NAT), FR-I type galaxy. But, the same jets could not be resolved at 235~MHz.

\section{Estimation of upper limits for the radio halos}\label{upper-limit}

We estimated the halo upper limits for clusters with no halos by grossly following the method of \citet{Bonafede_2017MNRAS}. The basic algorithm used here is described as follows.

\begin{enumerate}

\item First, we calculated the expected radio power from the cluster under consideration, making use of the correlation of $P_{1.4GHz}$ and  $M_{500}$ given in \citealt{Cassano_2013ApJ} i.e.,

\begin{equation}\label{eq:P-M}
\log\bigg(\frac{P_{1.4}}{10^{24.5} \; \rm{W \; Hz^{-1}}}\bigg) = B \; \log \bigg( \frac{M_{500}}{10^{14.9} \: \rm{M_{\odot}} }\bigg) + A
\end{equation}

where $A=0.125 \pm 0.076$ and $B=3.77\pm0.57$ are BCES-bisector fitting parameters for `Radio Halo' only data.

\item Expected size of the radio halo corresponding to the above radio power was then calculated from the scaling relation of 1.4~GHz radio power $P_{1.4GHz}$ and radio halo radii, $R_{H}$ \citep{Cassano_2007MNRAS} i.e.,
 
\begin{eqnarray}\label{eq:P-R}
\log\bigg(\frac{P_{1.4}}{5 \times 10^{24} \:  h^{-2}_{70}\: \rm{W \: Hz^{-1}}}\bigg) {}&=\; (4.18 \pm 0.68) \: \log \bigg( \frac{R_{H}}{500 h^{-1}_{70} \: \rm{kpc} }\bigg) \nonumber \\  
& - (0.26	\pm 0.07)
\end{eqnarray}

\item After calibrating the UV data using SPAM, data was taken into AIPS. A fake exponential halo was thereafter injected into the uv-plane with the expected radio power within the expected radius (as calculated above) using the task UVMOD. Further, the modified UV data was taken back to SPAM and was imaged. This was done to keep the method of imaging consistent throughout the study.

\item In this final image with mock halo, a halo is said to be barely detected if $D^{mock,meas}_{2\sigma} \gtrsim R_{H}$; where $D^{mock,meas}_{2\sigma}$ is the size of the halo measured above $2\sigma$. The corresponding flux ($S^{mock,meas}_{2\sigma}$)  i.e. the recovered flux should be at least 30\% of the injected flux to get the upper limit.

\item Now, if $D^{mock,meas}_{2\sigma} > R_{H}$ and $S^{mock,meas}_{2\sigma} \gg 30\% S^{mock,inj}_{R_{H}}$, then the value of flux to inject radio halo was decreased, and if $D^{mock,meas}_{2\sigma} < R_{H}$ then the value of flux to inject radio halo was increased. Steps (i) to (v), was repeated until we achieved the criteria given in point (iv).
\end{enumerate}

\subsection{Estimated radio halo upper-limits}\label{est:macs0152}

Adopting average slopes and intercept given in Eq.~\ref{eq:P-M} and Eq.~\ref{eq:P-R}, the estimated radio halo powers ($P_{1.4\rm{GHz}}^{exptd}$) for cluster MACS0152 and MACS0025 are $4.13\times 10^{24}~\rm{W~Hz^{-1}}$ and $5.3\times 10^{24}~\rm{W~Hz^{-1}}$. Corresponding expected radio halo sizes (from equation \ref{eq:P-R}) are 1108 and 1170 kpc respectively. Radio halo upper limit was thereafter evaluated performing the above mentioned steps. Calculations were done for images at 610 MHz and scaled 1.4~GHz upper limits of halo powers ($P_{1.4\rm{GHz}}^{inj}$) are $3.29\times 10^{24}~\rm{W~Hz^{-1}}$ and $4.8\times 10^{24}~\rm{W~Hz^{-1}}$ respectively.


\section{Discussions}\label{disc}

\subsection{Discovery of a diffuse relic-like structure in cluster MACS0152} 

We discovered a faint (total flux density $9.24\pm1.30$~mJy) and extended (about 0.5 Mpc), relic-like diffuse radio structure in cluster MACS0152 at GMRT 235~MHz. Unlike most of the arc-shaped radio relics that are observed almost at the virial radius of the host clusters, the projected location of this source is well inside the virial radius (see Fig.~\ref{MACSJ0152-opt-xray}, Panel 2). As reported in Section~\ref{macs0152-des}, this arc like radio structure is purely of diffuse in nature and is extended towards the outskirts of the cluster. The presence of a small group of galaxies in the NW, possibly falling into the bigger group towards SE, indicates an ongoing merger (Fig.~\ref{MACSJ0152-opt-xray}). So, the diffuse structure detected in this cluster may be a relic that has originated from this merger. The structure may also be a radio phoenix, the revived fossil electron clouds in the ICM, as indicated by its extremely steep spectrum $\alpha_{rel}=-2.3\pm0.6$ reported in this work. Unfortunately, the source could not be seen in the TGSS survey map at 150~MHz that would help us to get a better picture.

\subsection{Confirming a Relic in a high redshift cluster MACS0025}

Here, we reassure the detection of a very faint (flux density of $2.0\pm0.3$ mJy) radio relic structure in the NW side of the cluster  MACS0025 at GMRT 610 MHz. A re-analysed map at 325 MHz (see Fig.~\ref{mac0025-xray}) also reveals a similar structure. Figure~\ref{mac0025-xray} shows that the relic in cluster MACS0025 is placed at just outside the X-ray core of the cluster \citep{Brada2008ApJ} i.e., about 300 kpc or $45^{\prime\prime}$ from the centre. If, the relic has originated due to a merger shock, considering no projection effect and an average speed of 1000 km/s for the shock, merger would just be a 300 Myr old event. On the contrary, \citet{Ma_2010MNRAS} from their galaxy evolution study, indicate a relatively late merger phase (i.e. $0.5-1$ Gyr). Additionally, non-detection of a halo in our study as well as in a much deeper map of \citet{Riseley2017A&A} intrigues further investigation. A detection or non-detection of halo with even deeper map or at lower frequency observations with its detailed spectral and polarization properties would thus be interesting in connection to constraining possible particle acceleration mechanism and understanding the dynamical condition of the ICM. Nevertheless, detection of a radio relic in this cluster itself is a significant finding, as this distant cluster at redshift  $z=0.584$ is one of the few earliest (next only to El Gordo cluster; $z=0.87$; \citet{Lindner_2014ApJ}) merging systems detected with diffuse cluster radio emissions.

\subsection{Detection of a mini-halo in cluster MACS1931}

MACSJ1931 cluster as reported in \citep{Ehlert_2011MNRAS} shows evidence for both a recent merger and powerful AGN feedback indicating almost a relaxed phase, hosting a BCG with presence of a cool-core. The core got disturbed by both an AGN activity and sloshing of the low-entropy gas roughly along North-South direction. In the same study, two possible X-ray cavities are seen towards East and West of the BCG. The mass cooling rate within the central $50h_{70}^{-1}$ kpc is reported as  $700\rm{M_{\odot}}\rm{yr^{-1}}$. This is a perfect environment for harbouring a radio mini-halo. Indeed, a significant extra diffuse radio emission found surrounding the BCG at both observed frequencies. We detect a radio mini-halo with flux densities $651\pm65$ and $144\pm14$ at 235 and 610 MHz respectively, extending to about 300 kpc, much beyond the central BCG, and nicely following the X-ray radial profiles as reported in Section~\ref{MJ1931} and Figure~\ref{MiniHalo-jet}(a).

\subsection{Coexistent giant halo and relics in Abell 2744}

We present here the first low-frequency radio spectral properties of Abell 2744, a perfect radio halo-relic system through deep radio images from GMRT dual-band (610 \& 235~MHz) observations. For morphological signatures too, apart from 325~MHz at VLA \& GMRT \citep{Orr2007A&A,venturi2013}, no deeper radio study was done at low radio frequencies prior to this work (except our partly reported proceeding, \citealt{Paul_2014ASInC}). Though \cite{George_2017MNRAS} has worked with 88-200 MHz (GLEAM) and 150~MHz (TGSS-ADR1) images, both are from surveys, having inadequate resolution and sensitivity to accurately study the substructures and spectral properties.

\subsubsection{A giant, relatively flat-spectrum halo-relic system}

The giant relic~A in this cluster is a one-sided, peripheral, curved-relic with its concave side facing the central large halo and matching the halo curvature as observed in Fig.~\ref{A2744_comb}. This indicates a radially symmetric evolution. But, unlike the usual peripheral relics, which form almost along the merging axis (e.g. \citealt{Bagchi2006}), relic~A is found towards the perpendicular direction to the observed axis of current substructure merger (see Section~\ref{relAB-obs}). The dimension (LLS) of the prominent relic and the halo in this cluster is one of the largest known, with sizes of about 1.6~Mpc each at both 235 and 610~MHz. It is also important to notice that only a few clusters are known for hosting both a halo and a peripheral relic \citep{Giovannini2004,Weeren_2019SSRv} and Abell 2744 is a prominent example of that class.

\begin{table} 
\caption{Multi-band radio spectrum of Abell 2744} %
\centering          
\begin{tabular}{lrrr}   
\hline\hline    
Frequencies & \multicolumn{3}{c}{Spectral Index} \\
(MHz) & (Halo: &Relic-A: & Relic-B:) \\
\hline
118,154,200,\rdelim\}{2}{9pt}$^{(a)}$& \multirow{2}{*}{$-1.09\pm0.05$} & \multirow{2}{*}{$-1.01\pm0.07$}&\multirow{2}{*}{ --}\\
325 \& 1400 & & & \\
325 \& 1400$^{(b)}$ & $-1.0\pm0.1$ &
$-1.1\pm0.1$ & --\\
325 \& 1400$^{(c)}$& $-1.19^{+0.08}_{-0.11}$
& $-1.24\pm0.10$ & --\\  
235 \& 610$^{(e)}$ &  $-1.17\pm0.33$ &
$-1.14\pm0.36$ & $-1.65\pm0.62$ \\
235,325 \& 610 $^{(e)}$ & $-1.21\pm0.20$ & $-1.09\pm0.20$ & $-1.34\pm0.16$\\
610 \& 1500$^{(f)}$ & $-1.37\pm0.22$  & $-1.08\pm0.22$ &  --\\
1500 \& 3000$^{(d)}$ & $-1.43\pm0.11$ &
$-1.32\pm0.09$ & $-1.81\pm0.26$ \\ 
\hline
\multicolumn{4}{l}{ {\bf (a)}~\citet{George_2017MNRAS}; (GLEAM-TGSS-VLA)} \\
\multicolumn{4}{l}{{\bf (b)} \citet{Orr2007A&A}; (VLA-VLA)}\\ 

\multicolumn{4}{l}{ {\bf (c)} \citet{venturi2013}; (GMRT-VLA)}\\
\multicolumn{4}{l}{{\bf (d)} \citet{Pearce_2017ApJ} 
; (VLA-VLA)}\\

\multicolumn{4}{l}{{\bf (e)} Current study; (GMRT-GMRT)}\\
\multicolumn{4}{l}{{\bf (f)} Paul et. al., in prep.; (GMRT-VLA 1.5 GHz)}\\
\hline
\hline

\end{tabular}\label{tab:cluster-compare-2}
\end{table}

Average spectral indices of both relic~A and the halo in this cluster are strikingly similar i.e., ($\alpha_{relA}{^{235}_{610}}= -1.14\pm0.36$ and $\alpha_{halo}{^{235}_{610}} = -1.17\pm0.33$; see Section~\ref{Spec-ind1}) as found in this study. The trend remains the same at various frequencies as reported in studies with deeper observations (see reference $(c)-(e)$ in Table~\ref{tab:cluster-compare-2}). Spectral values in reference $(a)$ and $(b)$ of Table~\ref{tab:cluster-compare-2} are from low sensitive surveys and are inconsistent with the rest of the references (i.e. $(c)-(e)$). Interestingly, steepening of spectrum for both the halo and the relics is found insignificant in a wide range of frequencies (235-3000~MHz) i.e. almost no sign of spectral curvature. Figure~\ref{radio} shows that the prominent relic~A is about 1.7~Mpc away from the cluster centre. An average shock speed of about $1,000~\rm{km~s^{-1}}$ \citep{Sarazin_2002ASSL} would mean a merger responsible for the relic formation is older than 1.5~Gyr. Unless a continuous and in situ particle acceleration engine is present, it is unusual to observe a halo so bright at 1.4~GHz with insignificant spectral steepening (see Tab.~\ref{tab:spectra}) as per the known radio halo emission models. Synchrotron emission that remains visible only about a few hundred Mega years (i.e., equivalent to Synchrotron and IC cooling time (i.e. $10^{8}$ seconds); \citealt{Kang2017ApJ}), as well as time scale of turbulent energy (about 1 Gyr; \citealt{Cassano_2010ApJ}) would have ensured the fading away of the halo at high frequencies.

\subsubsection{Relic~B, a high speed second internal shock}
The second relic in Abell 2744 (Relic~B) is observed much closer to the centre and possibly linked to the merging brightest sub-cluster moving towards SE direction. It has a projected distance of only about $<0.7$~Mpc from this sub-clump. This shock has a Mach number of $\mathcal{M}=2.02\pm^{0.17}_{0.41}$ as computed from its injection spectral index of $\alpha_{inj}=1.15$ and corresponding error (see Sec.~\ref{Spec-ind1}) and using the relation
\begin{equation}\label{eq:Mach-comp}
\mathcal{M}^2 = \frac{2\alpha + 3}{2\alpha-1}
\end{equation} 
considering DSA mechanism \citep{Colafrancesco_2017MNRAS,Blandford_1987}.  This gives a shock speed of  $1769\pm^{148}_{359}~\rm{km~s^{-1}}$ in a medium with temperature of about $3.5\times10^7$~K (as seen in \citealt{Pearce_2017ApJ}). The shock velocity ($V_{sh}$) has been computed from the relation $V_{sh}=\mathcal{M}[1480(T_g/10^8 K)^{1/2}]\;\rm{km\;s^{-1}}$ \citep{Sarazin1988book} where $T_g$ is the temperature of the ICM gas. This closer, high-speed shock may have formed at a later time than that of the relic~A, indicating a second merger in the cluster. In fact, a similar two merger scenario (along EW and NS directions) has been proposed by \citet{Medezinski_2016ApJ}, as concluded from their weak lensing study.

\subsection{Fitting observed halos, mini-halos and relics into existing correlations}

\begin{figure*}
\includegraphics[width=8.8cm]{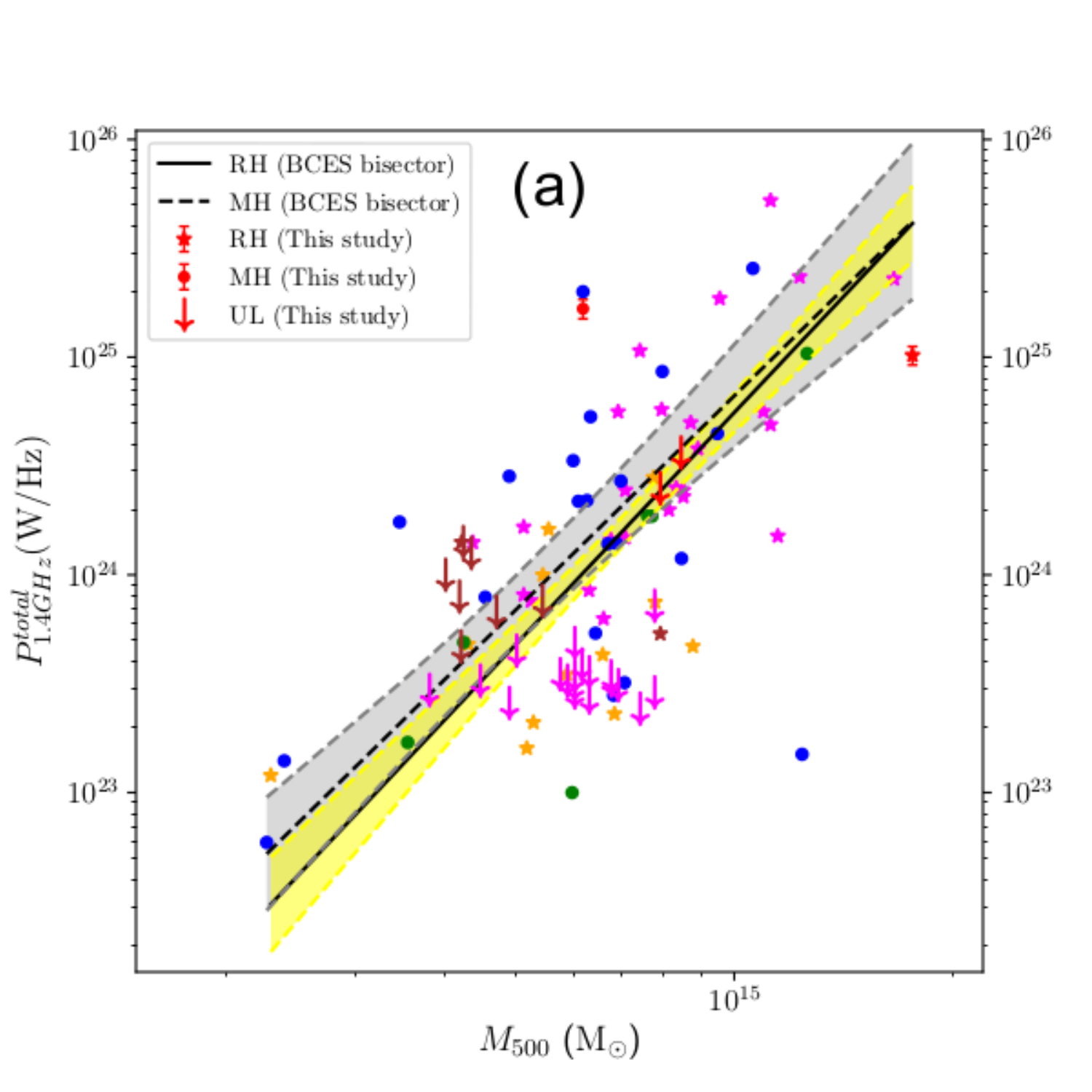} \hspace{-0.4cm}
\includegraphics[width=8.8cm]{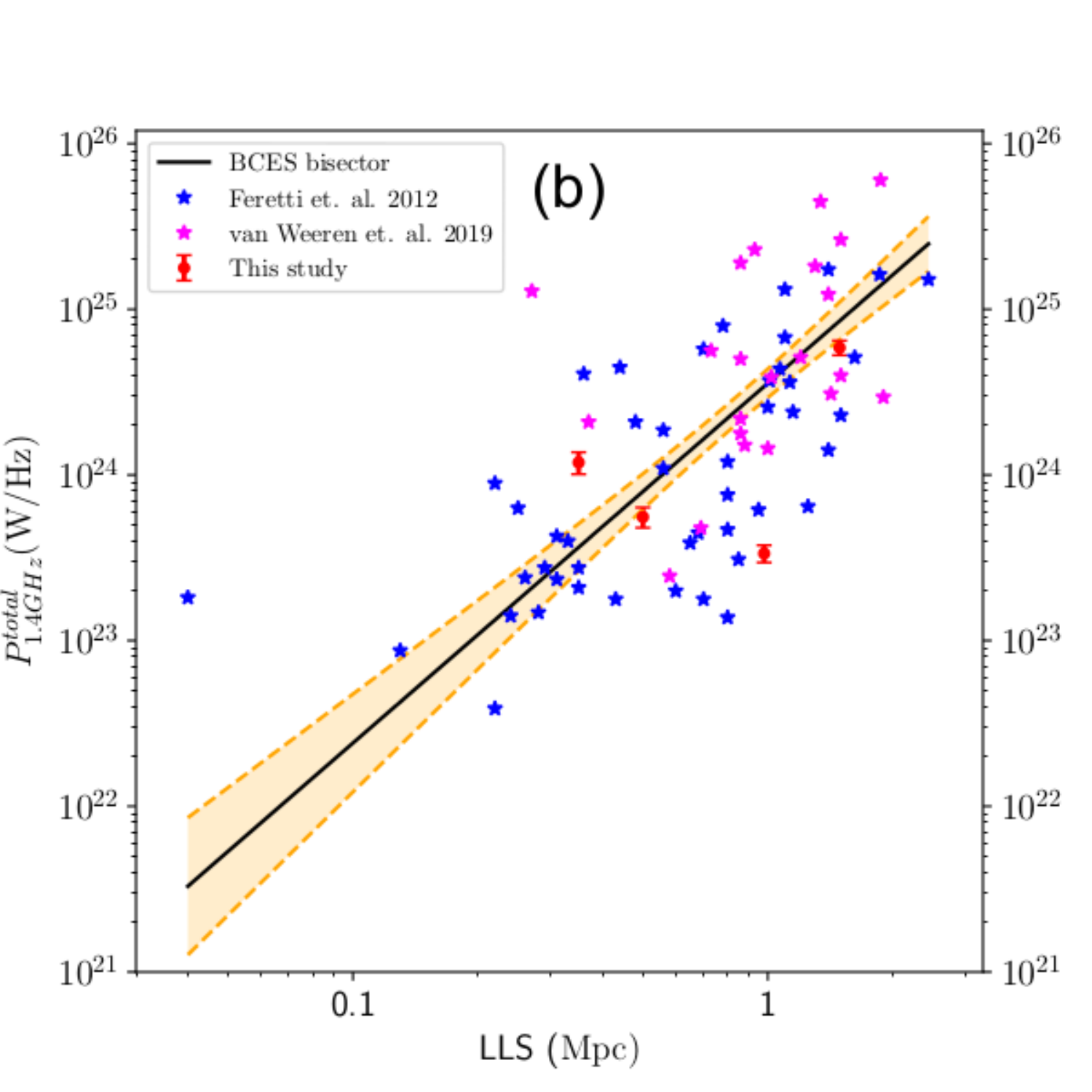}
\caption{{\bf Panel~$(a)$:} Power of radio halos (stars) and mini-halos (filled circles) and upper limits (down arrows) plotted ($P_{1.4 GHz}$) against virial mass ($M_{500}$). Data are from \citet{Cassano_2013ApJ} (magenta), \citet{Bernardi_2016MNRAS} (brown), \citet{Brizan_2019MNRAS} (orange), \citet{Giacintucci_2014ApJ} (blue) and \citet{Giacintucci_2017ApJ} (green). {\bf Panel~$(b)$:} Radio power against Largest Linear Size (LLS) of the relics. Data points are taken from \citet{Feretti_2012A&ARv} (blue) and \citet{Weeren_2019SSRv} (magenta). Data and upper limits from current study is plotted in red colour in both panels. BCES Bisector fitting slopes (black line, dashed) and confidence levels are drawn at 95\% i.e. about 2$\sigma$ (see Table~\ref{tab:correlation}).}\label{fig:p1.4_MXLLS}
\end{figure*}

Radio halo power (at 1.4 GHz) of massive clusters is known to follow a steep slope correlation (BCES bisector slope (all halos) $3.70\pm0.56$ and $2.11\pm0.20$) with cluster mass ($M_{500}$) and X-ray luminosity ($L_{500}$) respectively \citep{Cassano_2013ApJ}. A possible correlation between the relic power (at 1.4 GHz) and its size (LLS), has also been reported by \citet{Feretti_2012A&ARv}. So, it is useful to check, where our studied halos and relics stand with respect to these well-known correlations.

As reported in this study, in total, we detect one radio halo, a mini-halo and four radio relics. For correlation study, we projected our observed radio powers at 610/235 MHz to 1.4 GHz using the reported spectral index $\alpha^{1400}_{610/235}$ when it is available, and assumed $\alpha^{1400}_{610/235}=-1.3$ for the rest of the sources. Our estimated radio powers for cluster Abell 2744 are $P_{1.4}=1.16\pm0.12\times10^{25}\; \rm{W\;Hz^{-1}}$ (halo), $4.89\pm0.5\times10^{24}\; \rm{W\;Hz^{-1}}$ (relic~A) and $4.5\pm0.6\times10^{23}\;\rm{W\;Hz^{-1}}$ (relic~B), reasonably consistent to the values reported in \citet{Pearce_2017ApJ}. For the cluster MACS0025, the projected radio power of the relic detected at 610 MHz is $P_{1.4} = 1.19\pm0.18 \times 10^{24}\; \rm{W\; Hz}^{-1}$, again, consistent to the value reported in \citet{Riseley2017A&A}. We find the relic power in cluster MACS0152 to be $5.58\pm0.78\times10^{23}\; \rm{W\;Hz}^{-1}$. Our estimated radio halo upper limits at 1.4 GHz for the above two clusters are taken from Section~\ref{est:macs0152} and mini-halo power projected at 1.4 GHz is $1.67\pm0.17\times10^{25}\; \rm{W\;Hz^{-1}}$. A different halo upper limit obtained in cluster MACS0025 than \citet{Bonafede_2017MNRAS} may be attributed to assumed different slopes (in eq.~\ref{eq:P-M}) and the depth (rms) of the two maps at two different frequencies used in these studies.

\begin{table} 
\caption{Correlation slopes} %
\centering          
\begin{tabular}{llrr}   
\hline\hline    
Correlations & For & Old slope & New slope\\
\hline
$M_{500}$ vs $P_{1.4}$ & Radio Halo & $3.72\pm0.47$ & $3.54\pm0.43$ \\
\hline
$M_{500}$ vs $P_{1.4}$ & Radio Mini-halo & $3.20\pm0.59$ & $3.26\pm0.62$ \\
\hline
$LSS$ vs $P_{1.4}$ & Radio Relic &$2.19\pm0.26$ & $2.17\pm0.25$ \\
\hline
\hline
\end{tabular}\label{tab:correlation}
\end{table}

Figure~\ref{fig:p1.4_MXLLS}$(a)$, shows the correlation of mass of clusters (within radius $r_{500}$ i.e., $M_{500}$) versus the radio power of the halos and mini-halos separately at 1.4~GHz (i.e., $P_{1.4}$) and Figure~\ref{fig:p1.4_MXLLS}$(b)$ shows the plot of radio power versus size (LLS) of the relics. Data plotted from literature for the radio halos, mini-halos, relics and upper limits are as mentioned in the caption of Figure~\ref{fig:p1.4_MXLLS}. Data from the current study are plotted as red points with error bars, and upper limits are plotted as red down arrows. We computed the BCES bisector fitting slopes and confidence levels for already available data as well as combining with data from the current study. The old and the new slopes and their corresponding errors are reported in Table~\ref{tab:correlation}. New slopes are observed to be well within the errors of the old slopes.

\section{Conclusions}\label{conc}

\begin{itemize}

\item We report here the discovery of a radio relic-like diffuse emission in cluster MACS0152 and confirm the detection of a relic towards NW of the cluster MACS0025. These high-redshift clusters ($z=0.413$ \& $0.584$ respectively), hosting diffuse radio structures of sizes $\lesssim 0.5$ Mpc are possibly the evidence of early universe merging systems detected in radio waves. Further confirmation of types of radio sources demands deeper observations at lower frequencies as well as spectral and polarization properties.

\item This is the first report of the detection of a bright mini-halo of size 300 kpc in galaxy cluster MACS1931, both, at GMRT 235 and 610~MHz.

\item Presented deep radio study at GMRT 235/610 MHz reveals the first low-frequency spectral properties of a halo-relics system in cluster Abell 2744. This is one of the largest (1.6 Mpc each) known halo relic combinations at both low and high-frequency radio waves. This textbook-example halo observed in this cluster does not seem to have originated from the same event responsible for producing relic~A. Here, we also confirm the detection of a second relic, i.e., relic~B very near ($<0.7$~Mpc) to the cluster radio core. This relic is comparatively a high-Mach ($\mathcal{M}=2.02$) shock, with an estimated shock speed of $1769\pm^{148}_{359}~\rm{km~s^{-1}}$.

As reported, a bright halo with no significant steepening of the spectrum at high frequencies, a further, and a closer high-speed relic placed almost orthogonally, and a proposed two merger scenario from weak lensing studies may have a strong connection that needs to be investigated further. This may provide a clue to understanding the mystery of the origin of these unique radio structures present in this cluster.

\item All detected halo, mini-halo and relics reported in this study shown to fit well with the prevailing radio power vs mass, X-ray luminosity as well as LLS correlations, therefore it seems, they are a good addition to the existing data set of cluster radio sources. This is to mention that the data used here are from various telescopes, imaging parameters and measuring methods i.e. it lacks homogeneity and may well affect the correlation calculations. Therefore demands a thorough and careful study to derive a meaningful physical connection among the parameters.

\end{itemize}

\section*{Acknowledgements}
We would like to thank the staff of the GMRT that made these observations possible. GMRT is run by the National Centre for Radio Astrophysics of the Tata Institute of Fundamental Research. SP acknowledges DST-SERB Fast Track scheme for young scientists, Grant No. SR/FTP/PS-118/2011 for supporting this research. He also wants to thank DST INSPIRE Faculty Scheme
(IF12/PH-44) for funding his research group. Part of this work was done when AD was a NASA post-doc at CASA, Univ. of Colorado, Boulder USA. Authors would like to thank the editor and the anonymous referees for their constructive comments that helped in improving the quality of the manuscript.


\label{lastpage}


\begin{thebibliography}{}
\bibitem[Bagchi et al.(2006)]{Bagchi2006} {Bagchi}, J., {Durret}, F., {Neto}, G.~B.~L., \& {Paul}, S. 2006, {\it Science}, 314, 791 

\bibitem[Bernardi et al.(2016)]{Bernardi_2016MNRAS} Bernardi, G., Venturi, T., Cassano, R., et al.\ 2016, \mnras, 456, 1259 

\bibitem[Blandford et al.(1987)]{Blandford_1987}
Blandford, R., \& Eichler, D. 1987, {\it Physics Report}, 154, 1 

\bibitem[Bonafede et al.(2017)]{Bonafede_2017MNRAS} Bonafede, A., Cassano, R., Br{\"u}ggen, M., et al.\ 2017, \mnras, 470, 3465  

\bibitem[Brada{\v c} et al.(2008)]{Brada2008ApJ} Brada{\v c}, M., Allen, S.~W., Treu, T., et al.\ 2008, \apj, 687, 959 

\bibitem[B{\^\i}rzan et al.(2019)]{Brizan_2019MNRAS} B{\^\i}rzan, L., Rafferty, D.~A., Cassano, R., et al.\ 2019, \mnras, 1394

\bibitem[Brunetti et al.(2001)]{Brunetti_2001MNRAS} Brunetti, G., Setti, G., Feretti, L., \& Giovannini, G.\ 2001, \mnras, 320, 365 

\bibitem[Bykov et al.(2008)]{bykov2008}
Bykov, A.~M., Dolag, K. and Durret, F., 2008, {\it SSRv.}, 134, 119

\bibitem[Cassano et al.(2007)]{Cassano_2007MNRAS} Cassano, R., Brunetti, G., Setti, G., Govoni, F., \& Dolag, K.\ 2007, \mnras, 378, 1565

\bibitem[Cassano et al.(2010)]{Cassano_2010ApJ} 
Cassano, R., Ettori, S., Giacintucci, S., et al.\ 2010, \apjl, 721, L82

\bibitem[Cassano et al.(2011)]{Cassano2011} 
{Cassano}, R., {Brunetti}, G., \& {Venturi}, T. 2011, {\it Journal of Astrophysics and Astronomy}, 32, 519 

\bibitem[Cassano et al.(2013)]{Cassano_2013ApJ} Cassano, R., Ettori, S., Brunetti, G., et al.\ 2013, \apj, 777, 141

\bibitem[Cassano et al.(2016)]{Cassano_2016A&A} 
Cassano, R., Brunetti, G., Giocoli, C., \& Ettori, S.\ 2016, \aap, 593, A81

\bibitem[Colafrancesco et al.(2017)]{Colafrancesco_2017MNRAS} Colafrancesco, S., Marchegiani, P., \& Paulo, C.~M.\ 2017, \mnras, 471, 4747 

\bibitem[Cuciti et al.(2015)]{Cuciti_2015A&A} Cuciti, V., Cassano, R., Brunetti, G., et al.\ 2015, \aap, 580, A97

\bibitem[Donnert et al.(2013)]{Donnert_2013MNRAS} 
Donnert, J., Dolag, K., Brunetti, G., \& Cassano, R.\ 2013, \mnras, 429, 3564

\bibitem[Ebeling et al.(2001)]{Ebeling2001} 
 {Ebeling} H., {Edge} A.~C. and {Henry} J.~P., 2001, {\it ApJ}, 553, 668

\bibitem[Ebeling et al.(2007)]{Ebeling_2007ApJ} Ebeling, H., Barrett, E., Donovan, D., et al.\ 2007, \apjl, 661, L33 

\bibitem[Ebeling et al.(2010)]{Ebeling2010} 
 {Ebeling}, H., {Edge}, A.~C., {Mantz}, A., et al. 2010, {\it MNRAS}, 407, 83

\bibitem[Ehlert et al.(2011)]{Ehlert_2011MNRAS} Ehlert, S., Allen, S.~W., von der Linden, A., et al.\ 2011, \mnras, 411, 1641.

\bibitem[Ehlert et al.(2013)]{Ehlert_2013MNRAS} Ehlert, S., Allen, S.~W., Brandt, W.~N., et al.\ 2013, \mnras, 428, 3509.

\bibitem[Ensslin et al.(1998)]{Ensslin1998} 
{Ensslin}, T.~A., {Biermann}, P.~L., {Klein}, U., \& {Kohle}, S.\ 1998, \aap, 332, 395 

\bibitem[Feretti et al.(2001)]{Feretti_2001A&A} 
Feretti, L., Fusco-Femiano, R., Giovannini, G., \& Govoni, F.\ 2001, \aap, 373, 106

\bibitem[Feretti \& Giovannini(2008)]{Feretti_2008LNP} Feretti, L., \& Giovannini, G.\ 2008, A Pan-Chromatic View of Clusters of Galaxies and the Large-Scale Structure, 740, 24 

\bibitem[Feretti et al.(2012)]{Feretti_2012A&ARv} 
 Feretti, L., Giovannini, G., Govoni, F., \& Murgia, M.\ 2012, \aapr, 20, 54

\bibitem[George et al.(2017)]{George_2017MNRAS} George, L.~T., Dwarakanath, K.~S., Johnston-Hollitt, M., et al.\ 2017, \mnras, 467, 936 

\bibitem[Giacintucci, et al.(2014)]{Giacintucci_2014ApJ} Giacintucci, S., Markevitch, M., Venturi, T., et al.\ 2014, \apj, 781, 9.

\bibitem[Giacintucci et al.(2017)]{Giacintucci_2017ApJ} Giacintucci, S., Markevitch, M., Cassano, R., et al.\ 2017, \apj, 841, 71

\bibitem[Giovannini et al.,(1999)]{Giovannini1999} 
{Giovannini}, G., {Tordi}, M., \& {Feretti}, L. 1999, {\it New Astronomy}, 4, 141

\bibitem[Giovannini \& Feretti (2004)]{Giovannini2004} 
{Giovannini} G., \& Feretti, L. 2004, {\it JKAS}, 37, 323

\bibitem[Giovannini et al.(2009)]{Giovannini_2009A&A} Giovannini, G., Bonafede, A., Feretti, L., et al.\ 2009, \aap, 507, 1257

\bibitem[Gitti et al.(2018)]{Gitti_2018A&A} Gitti, M., Brunetti, G., Cassano, R., et al.\ 2018, \aap, 617, A11


\bibitem[Govoni et al.(2001a)]{Govoni2001A&A} Govoni, F., En{\ss}lin, T.~A., Feretti, L., \& Giovannini, G.\ 2001, \aap, 369, 441 
 
\bibitem[Govoni et al.(2001b)]{Govoni2001} 
{Govoni}, F., {Feretti}, L., {Giovannini,} G., et al. 2001, \aap, 376, 803 

\bibitem[Horesh et al.(2010)]{Horesh2010}
 {Horesh} A., {Maoz} D., {Ebeling} H., {Seidel} G., and {Bartelmann} M., 2010, {\it MNRAS}, 406, 1318

\bibitem[Iapichino \& Br{\"u}ggen(2012)]{luigi2012}
{Iapichino} L., \& {Br{\"u}ggen} M. 2012, {\it MNRAS}, 423, 2781

\bibitem[Intema et al.(2017)]{Intema_2017A&A} Intema, H.~T., Jagannathan, P., Mooley, K.~P., \& Frail, D.~A.\ 2017, \aap, 598, A78

\bibitem[Kang et al.(2017)]{Kang2017ApJ} Kang, H., Ryu, D., \& Jones, T.~W.\ 2017, \apj, 840, 42 

\bibitem[Kim \& Trippe(2014)]{Kim_2014JKAS} Kim, J.-Y., \& Trippe, S.\ 2014, Journal of Korean Astronomical Society, 47, 195 

\bibitem[Lindner et al.(2014)]{Lindner_2014ApJ} Lindner, R.~R., Baker, A.~J., Hughes, J.~P., et al.\ 2014, \apj, 786, 49

\bibitem[Liu et al.(2018)]{Liu_2018MNRAS} Liu, A., Tozzi, P., Yu, H., et al.\ 2018, \mnras, 481, 361. 

\bibitem[Ma et al.(2010)]{Ma_2010MNRAS} Ma, C.-J., Ebeling, H., Marshall, P., \& Schrabback, T.\ 2010, \mnras, 406, 121 

\bibitem[Mantz et al.(2010)]{Mantz_2010MNRAS} Mantz, A., Allen, S.~W., Ebeling, H., Rapetti, D., \& Drlica-Wagner, A.\ 2010, \mnras, 406, 1773

\bibitem[McMullin et al.(2007)]{McMullin_2007ASPC} McMullin, J.~P., Waters, B., Schiebel, D., Young, W., \& Golap, K.\ 2007, Astronomical Data Analysis Software and Systems XVI, 376, 127

\bibitem[Medezinski, et al.(2016)]{Medezinski_2016ApJ} Medezinski, E., Umetsu, K., Okabe, N., et al.\ 2016, \apj, 817, 24. 

\bibitem[Orr{\'u} et al.(2007)]{Orr2007A&A} 
 Orr{\'u}, E., Murgia, M., Feretti, L., et al.\ 2007, \aap, 467, 943

\bibitem[Rephaeli et al.(2008)]{Rephaeli2008} 
{Rephaeli}, Y., {Nevalainen}, J., {Ohashi}, T., \& {Bykov}, A.~M. 2008, {\it SSRv}, 134, 71

\bibitem[Riseley et al.(2017)]{Riseley2017A&A} Riseley, C.~J., Scaife, A.~M.~M., Wise, M.~W., \& Clarke, A.~O.\ 2017, \aap, 597, A96 
 
\bibitem[Paul et al.(2014)]{Paul_2014ASInC} Paul, S., Datta, A., \& Intema, H.~T.\ 2014, Astronomical Society of India Conference Series, 13, 187 

\bibitem[Paul et al.(2018)]{Paul_2018arXiv} Paul, S., Gupta, P., John, R.~S., \& Pubjabi, V.\ 2018, arXiv:1803.10764 

\bibitem[Pearce et al.(2017)]{Pearce_2017ApJ} Pearce, C.~J.~J., van Weeren, R.~J., Andrade-Santos, F., et al.\ 2017, \apj, 845, 81 

\bibitem[Pinzke et al.(2017)]{Pinzke2017MNRAS} Pinzke, A., Oh, S.~P., \& Pfrommer, C.\ 2017, \mnras, 465, 4800 

\bibitem[Sarazin(1986)]{Sarazin1986}
 {Sarazin}, Craig L., 1986,  {\it Rev. Mod. Phys.}, 58, 1

\bibitem[Sarazin(1988)]{Sarazin1988book} Sarazin, C.~L.\ 1988, Cambridge Astrophysics Series, Cambridge: Cambridge University Press, 1988,  

\bibitem[Sarazin(2002)]{Sarazin_2002ASSL} Sarazin, C.~L.\ 2002, Merging Processes in Galaxy Clusters, 272, 1 

\bibitem[Paul et al.(2011)]{Paul2011} 
  S. Paul, L. Iapichino, F. Miniati, J. Bagchi \& K. Mannheim 2011, {\it ApJ}, 726, 17 
  
 \bibitem[Paul(2012)]{Paul2012} 
  S. Paul, 2012, {\it Journal of Physics: Conference Series}, 405, 012026

\bibitem[Stroe et al.(2013)]{Stroe_2013A&A} Stroe, A., van Weeren, R.~J., Intema, H.~T., et al.\ 2013, \aap, 555, A110 

\bibitem[Stroe, et al.(2014)]{Stroe_2014MNRAS} Stroe, A., Rumsey, C., Harwood, J.~J., et al.\ 2014, \mnras, 441, L41.

\bibitem[Subramanian et al.(2006)]{Subra2006} Subramanian, K., Shukurov, A., \& Haugen, N. E. L. 2006, {\it MNRAS}, 366, 1437
 
 \bibitem[van Weeren et al.(2009)]{Weeren_2009A&A}  van Weeren, R.~J., R{\"o}ttgering, H.~J.~A., Bagchi, J., et al.\ 2009, {\it A\&A}, 506, 1083

\bibitem[van Weeren et al.(2010)]{Weeren_2010Sci} van Weeren, R.~J., R{\"o}ttgering, H.~J.~A., Br{\"u}ggen, M., \& Hoeft, M.\ 2010, Science, 330, 347 

\bibitem[van Weeren et al.(2014)]{Weeren_2014ApJ} van Weeren, R.~J., Intema, H.~T., Lal, D.~V., et al.\ 2014, \apj, 786, L17

\bibitem[van Weeren et al.(2019)]{Weeren_2019SSRv} van Weeren, R.~J., de Gasperin, F., Akamatsu, H., et al.\ 2019, \ssr, 215, 16.
 
 \bibitem[Venturi et al.(2013)]{venturi2013} 
{Venturi} T., {Giacintucci} S., {Dallacasa} D., et al. 2013, \aap, 551, A24

\bibitem[Wayth et al.(2015)]{Wayth_2015PASA} Wayth, R.~B., Lenc, E., Bell, M.~E., et al.\ 2015, \pasa, 32, e025 

\bibitem[Zitrin et al. (2011)]{Zitrin2011}
 {Zitrin} A., {Broadhurst} T., {Barkana} R., {Rephaeli} Y., and {Ben{\'{\i}}tez} N., 2011, {\it MNRAS}, 410, 1939
 
\end{thebibliography}
\end{document}